\documentclass[12pt,a4paper]{article}
\usepackage[a4paper, hmargin=2cm, vmargin=2cm]{geometry}
\usepackage{authblk}

\usepackage{amssymb}
\usepackage{amsmath}
\usepackage{stmaryrd}       
\usepackage{mathcomp}
\usepackage{color}
\usepackage{cite}
\usepackage{graphicx}
\usepackage{setspace}
\usepackage[colorinlistoftodos,prependcaption,textsize=footnotesize]{todonotes}
\usepackage{verbatim}		
\usepackage{courier}
\usepackage{mdframed}
\usepackage{longtable}
\usepackage{multirow}
\usepackage{caption}        
\usepackage{subcaption}
\usepackage{commath}        
\usepackage{epstopdf}
\usepackage{float}
\usepackage{bm}
\usepackage{textpos}
\usepackage{url}            
\usepackage[english]{babel}
\usepackage{dsfont}
\usepackage{array}
\usepackage{latexsym}
\usepackage{mathtools}
\usepackage{afterpage}
\usepackage{algorithm}
\usepackage{algorithmic}
\usepackage[pdftex,colorlinks]{hyperref}
\usepackage[capitalise]{cleveref}
\usepackage[all]{hypcap}
\usepackage{soul}
\usepackage{xargs}          
\usepackage{keyval}         
\usepackage{pgfkeys}        
\usepackage{fix-cm}
\usepackage{blindtext}      
\usepackage{pifont}         

\usepackage{tikz}

\usepackage{tabularx}       
\usepackage{tcolorbox}      
\usepackage{colortbl}       
\usepackage{array}          

\usepackage{lipsum}  
\usepackage{enumitem}       

\newcommand{\notshow}[1]{}

\newcommand{\mat}[1]{\mathbf{#1}}                       
\newcommand{\ten}[1]{\mathbf{\underline{#1}}}           

\newcommand{\td}[1]{\llbracket #1 \rrbracket}
\newcommand{\R}{\mathbb{R}}

\definecolor{AliceBlue}{rgb}{0.94,0.97,1.00}
\definecolor{AntiqueWhite1}{rgb}{1.00,0.94,0.86}
\definecolor{AntiqueWhite2}{rgb}{0.93,0.87,0.80}
\definecolor{AntiqueWhite3}{rgb}{0.80,0.75,0.69}
\definecolor{AntiqueWhite4}{rgb}{0.55,0.51,0.47}
\definecolor{AntiqueWhite}{rgb}{0.98,0.92,0.84}
\definecolor{BlanchedAlmond}{rgb}{1.00,0.92,0.80}
\definecolor{BlueViolet}{rgb}{0.54,0.17,0.89}
\definecolor{CadetBlue1}{rgb}{0.60,0.96,1.00}
\definecolor{CadetBlue2}{rgb}{0.56,0.90,0.93}
\definecolor{CadetBlue3}{rgb}{0.48,0.77,0.80}
\definecolor{CadetBlue4}{rgb}{0.33,0.53,0.55}
\definecolor{CadetBlue}{rgb}{0.37,0.62,0.63}
\definecolor{CornflowerBlue}{rgb}{0.39,0.58,0.93}
\definecolor{DarkBlue}{rgb}{0.00,0.00,0.55}
\definecolor{DarkCyan}{rgb}{0.00,0.55,0.55}
\definecolor{DarkGoldenrod1}{rgb}{1.00,0.73,0.06}
\definecolor{DarkGoldenrod2}{rgb}{0.93,0.68,0.05}
\definecolor{DarkGoldenrod3}{rgb}{0.80,0.58,0.05}
\definecolor{DarkGoldenrod4}{rgb}{0.55,0.40,0.03}
\definecolor{DarkGoldenrod}{rgb}{0.72,0.53,0.04}
\definecolor{DarkGray}{rgb}{0.66,0.66,0.66}
\definecolor{DarkGreen}{rgb}{0.00,0.39,0.00}
\definecolor{DarkGrey}{rgb}{0.66,0.66,0.66}
\definecolor{DarkKhaki}{rgb}{0.74,0.72,0.42}
\definecolor{DarkMagenta}{rgb}{0.55,0.00,0.55}
\definecolor{DarkOliveGreen1}{rgb}{0.79,1.00,0.44}
\definecolor{DarkOliveGreen2}{rgb}{0.74,0.93,0.41}
\definecolor{DarkOliveGreen3}{rgb}{0.64,0.80,0.35}
\definecolor{DarkOliveGreen4}{rgb}{0.43,0.55,0.24}
\definecolor{DarkOliveGreen}{rgb}{0.33,0.42,0.18}
\definecolor{DarkOrange1}{rgb}{1.00,0.50,0.00}
\definecolor{DarkOrange2}{rgb}{0.93,0.46,0.00}
\definecolor{DarkOrange3}{rgb}{0.80,0.40,0.00}
\definecolor{DarkOrange4}{rgb}{0.55,0.27,0.00}
\definecolor{DarkOrange}{rgb}{1.00,0.55,0.00}
\definecolor{DarkOrchid1}{rgb}{0.75,0.24,1.00}
\definecolor{DarkOrchid2}{rgb}{0.70,0.23,0.93}
\definecolor{DarkOrchid3}{rgb}{0.60,0.20,0.80}
\definecolor{DarkOrchid4}{rgb}{0.41,0.13,0.55}
\definecolor{DarkOrchid}{rgb}{0.60,0.20,0.80}
\definecolor{DarkRed}{rgb}{0.55,0.00,0.00}
\definecolor{DarkSalmon}{rgb}{0.91,0.59,0.48}
\definecolor{DarkSeaGreen1}{rgb}{0.76,1.00,0.76}
\definecolor{DarkSeaGreen2}{rgb}{0.71,0.93,0.71}
\definecolor{DarkSeaGreen3}{rgb}{0.61,0.80,0.61}
\definecolor{DarkSeaGreen4}{rgb}{0.41,0.55,0.41}
\definecolor{DarkSeaGreen}{rgb}{0.56,0.74,0.56}
\definecolor{DarkSlateBlue}{rgb}{0.28,0.24,0.55}
\definecolor{DarkSlateGray1}{rgb}{0.59,1.00,1.00}
\definecolor{DarkSlateGray2}{rgb}{0.55,0.93,0.93}
\definecolor{DarkSlateGray3}{rgb}{0.47,0.80,0.80}
\definecolor{DarkSlateGray4}{rgb}{0.32,0.55,0.55}
\definecolor{DarkSlateGray}{rgb}{0.18,0.31,0.31}
\definecolor{DarkSlateGrey}{rgb}{0.18,0.31,0.31}
\definecolor{DarkTurquoise}{rgb}{0.00,0.81,0.82}
\definecolor{DarkViolet}{rgb}{0.58,0.00,0.83}
\definecolor{DeepPink1}{rgb}{1.00,0.08,0.58}
\definecolor{DeepPink2}{rgb}{0.93,0.07,0.54}
\definecolor{DeepPink3}{rgb}{0.80,0.06,0.46}
\definecolor{DeepPink4}{rgb}{0.55,0.04,0.31}
\definecolor{DeepPink}{rgb}{1.00,0.08,0.58}
\definecolor{DeepSkyBlue1}{rgb}{0.00,0.75,1.00}
\definecolor{DeepSkyBlue2}{rgb}{0.00,0.70,0.93}
\definecolor{DeepSkyBlue3}{rgb}{0.00,0.60,0.80}
\definecolor{DeepSkyBlue4}{rgb}{0.00,0.41,0.55}
\definecolor{DeepSkyBlue}{rgb}{0.00,0.75,1.00}
\definecolor{DimGray}{rgb}{0.41,0.41,0.41}
\definecolor{DimGrey}{rgb}{0.41,0.41,0.41}
\definecolor{DodgerBlue1}{rgb}{0.12,0.56,1.00}
\definecolor{DodgerBlue2}{rgb}{0.11,0.53,0.93}
\definecolor{DodgerBlue3}{rgb}{0.09,0.45,0.80}
\definecolor{DodgerBlue4}{rgb}{0.06,0.31,0.55}
\definecolor{DodgerBlue}{rgb}{0.12,0.56,1.00}
\definecolor{FloralWhite}{rgb}{1.00,0.98,0.94}
\definecolor{ForestGreen}{rgb}{0.13,0.55,0.13}
\definecolor{GhostWhite}{rgb}{0.97,0.97,1.00}
\definecolor{GreenYellow}{rgb}{0.68,1.00,0.18}
\definecolor{HotPink1}{rgb}{1.00,0.43,0.71}
\definecolor{HotPink2}{rgb}{0.93,0.42,0.65}
\definecolor{HotPink3}{rgb}{0.80,0.38,0.56}
\definecolor{HotPink4}{rgb}{0.55,0.23,0.38}
\definecolor{HotPink}{rgb}{1.00,0.41,0.71}
\definecolor{IndianRed1}{rgb}{1.00,0.42,0.42}
\definecolor{IndianRed2}{rgb}{0.93,0.39,0.39}
\definecolor{IndianRed3}{rgb}{0.80,0.33,0.33}
\definecolor{IndianRed4}{rgb}{0.55,0.23,0.23}
\definecolor{IndianRed}{rgb}{0.80,0.36,0.36}
\definecolor{LavenderBlush1}{rgb}{1.00,0.94,0.96}
\definecolor{LavenderBlush2}{rgb}{0.93,0.88,0.90}
\definecolor{LavenderBlush3}{rgb}{0.80,0.76,0.77}
\definecolor{LavenderBlush4}{rgb}{0.55,0.51,0.53}
\definecolor{LavenderBlush}{rgb}{1.00,0.94,0.96}
\definecolor{LawnGreen}{rgb}{0.49,0.99,0.00}
\definecolor{LemonChiffon1}{rgb}{1.00,0.98,0.80}
\definecolor{LemonChiffon2}{rgb}{0.93,0.91,0.75}
\definecolor{LemonChiffon3}{rgb}{0.80,0.79,0.65}
\definecolor{LemonChiffon4}{rgb}{0.55,0.54,0.44}
\definecolor{LemonChiffon}{rgb}{1.00,0.98,0.80}
\definecolor{LightBlue1}{rgb}{0.75,0.94,1.00}
\definecolor{LightBlue2}{rgb}{0.70,0.87,0.93}
\definecolor{LightBlue3}{rgb}{0.60,0.75,0.80}
\definecolor{LightBlue4}{rgb}{0.41,0.51,0.55}
\definecolor{LightBlue}{rgb}{0.68,0.85,0.90}
\definecolor{LightCoral}{rgb}{0.94,0.50,0.50}
\definecolor{LightCyan1}{rgb}{0.88,1.00,1.00}
\definecolor{LightCyan2}{rgb}{0.82,0.93,0.93}
\definecolor{LightCyan3}{rgb}{0.71,0.80,0.80}
\definecolor{LightCyan4}{rgb}{0.48,0.55,0.55}
\definecolor{LightCyan}{rgb}{0.88,1.00,1.00}
\definecolor{LightGoldenrod1}{rgb}{1.00,0.93,0.55}
\definecolor{LightGoldenrod2}{rgb}{0.93,0.86,0.51}
\definecolor{LightGoldenrod3}{rgb}{0.80,0.75,0.44}
\definecolor{LightGoldenrod4}{rgb}{0.55,0.51,0.30}
\definecolor{LightGoldenrodYellow}{rgb}{0.98,0.98,0.82}
\definecolor{LightGoldenrod}{rgb}{0.93,0.87,0.51}
\definecolor{LightGray}{rgb}{0.83,0.83,0.83}
\definecolor{LightGreen}{rgb}{0.56,0.93,0.56}
\definecolor{LightGrey}{rgb}{0.83,0.83,0.83}
\definecolor{LightPink1}{rgb}{1.00,0.68,0.73}
\definecolor{LightPink2}{rgb}{0.93,0.64,0.68}
\definecolor{LightPink3}{rgb}{0.80,0.55,0.58}
\definecolor{LightPink4}{rgb}{0.55,0.37,0.40}
\definecolor{LightPink}{rgb}{1.00,0.71,0.76}
\definecolor{LightSalmon1}{rgb}{1.00,0.63,0.48}
\definecolor{LightSalmon2}{rgb}{0.93,0.58,0.45}
\definecolor{LightSalmon3}{rgb}{0.80,0.51,0.38}
\definecolor{LightSalmon4}{rgb}{0.55,0.34,0.26}
\definecolor{LightSalmon}{rgb}{1.00,0.63,0.48}
\definecolor{LightSeaGreen}{rgb}{0.13,0.70,0.67}
\definecolor{LightSkyBlue1}{rgb}{0.69,0.89,1.00}
\definecolor{LightSkyBlue2}{rgb}{0.64,0.83,0.93}
\definecolor{LightSkyBlue3}{rgb}{0.55,0.71,0.80}
\definecolor{LightSkyBlue4}{rgb}{0.38,0.48,0.55}
\definecolor{LightSkyBlue}{rgb}{0.53,0.81,0.98}
\definecolor{LightSlateBlue}{rgb}{0.52,0.44,1.00}
\definecolor{LightSlateGray}{rgb}{0.47,0.53,0.60}
\definecolor{LightSlateGrey}{rgb}{0.47,0.53,0.60}
\definecolor{LightSteelBlue1}{rgb}{0.79,0.88,1.00}
\definecolor{LightSteelBlue2}{rgb}{0.74,0.82,0.93}
\definecolor{LightSteelBlue3}{rgb}{0.64,0.71,0.80}
\definecolor{LightSteelBlue4}{rgb}{0.43,0.48,0.55}
\definecolor{LightSteelBlue}{rgb}{0.69,0.77,0.87}
\definecolor{LightYellow1}{rgb}{1.00,1.00,0.88}
\definecolor{LightYellow2}{rgb}{0.93,0.93,0.82}
\definecolor{LightYellow3}{rgb}{0.80,0.80,0.71}
\definecolor{LightYellow4}{rgb}{0.55,0.55,0.48}
\definecolor{LightYellow}{rgb}{1.00,1.00,0.88}
\definecolor{LimeGreen}{rgb}{0.20,0.80,0.20}
\definecolor{MediumAquamarine}{rgb}{0.40,0.80,0.67}
\definecolor{MediumBlue}{rgb}{0.00,0.00,0.80}
\definecolor{MediumOrchid1}{rgb}{0.88,0.40,1.00}
\definecolor{MediumOrchid2}{rgb}{0.82,0.37,0.93}
\definecolor{MediumOrchid3}{rgb}{0.71,0.32,0.80}
\definecolor{MediumOrchid4}{rgb}{0.48,0.22,0.55}
\definecolor{MediumOrchid}{rgb}{0.73,0.33,0.83}
\definecolor{MediumPurple1}{rgb}{0.67,0.51,1.00}
\definecolor{MediumPurple2}{rgb}{0.62,0.47,0.93}
\definecolor{MediumPurple3}{rgb}{0.54,0.41,0.80}
\definecolor{MediumPurple4}{rgb}{0.36,0.28,0.55}
\definecolor{MediumPurple}{rgb}{0.58,0.44,0.86}
\definecolor{MediumSeaGreen}{rgb}{0.24,0.70,0.44}
\definecolor{MediumSlateBlue}{rgb}{0.48,0.41,0.93}
\definecolor{MediumSpringGreen}{rgb}{0.00,0.98,0.60}
\definecolor{MediumTurquoise}{rgb}{0.28,0.82,0.80}
\definecolor{MediumVioletRed}{rgb}{0.78,0.08,0.52}
\definecolor{MidnightBlue}{rgb}{0.10,0.10,0.44}
\definecolor{MintCream}{rgb}{0.96,1.00,0.98}
\definecolor{MistyRose1}{rgb}{1.00,0.89,0.88}
\definecolor{MistyRose2}{rgb}{0.93,0.84,0.82}
\definecolor{MistyRose3}{rgb}{0.80,0.72,0.71}
\definecolor{MistyRose4}{rgb}{0.55,0.49,0.48}
\definecolor{MistyRose}{rgb}{1.00,0.89,0.88}
\definecolor{NavajoWhite1}{rgb}{1.00,0.87,0.68}
\definecolor{NavajoWhite2}{rgb}{0.93,0.81,0.63}
\definecolor{NavajoWhite3}{rgb}{0.80,0.70,0.55}
\definecolor{NavajoWhite4}{rgb}{0.55,0.47,0.37}
\definecolor{NavajoWhite}{rgb}{1.00,0.87,0.68}
\definecolor{NavyBlue}{rgb}{0.00,0.00,0.50}
\definecolor{OldLace}{rgb}{0.99,0.96,0.90}
\definecolor{OliveDrab1}{rgb}{0.75,1.00,0.24}
\definecolor{OliveDrab2}{rgb}{0.70,0.93,0.23}
\definecolor{OliveDrab3}{rgb}{0.60,0.80,0.20}
\definecolor{OliveDrab4}{rgb}{0.41,0.55,0.13}
\definecolor{OliveDrab}{rgb}{0.42,0.56,0.14}
\definecolor{OrangeRed1}{rgb}{1.00,0.27,0.00}
\definecolor{OrangeRed2}{rgb}{0.93,0.25,0.00}
\definecolor{OrangeRed3}{rgb}{0.80,0.22,0.00}
\definecolor{OrangeRed4}{rgb}{0.55,0.15,0.00}
\definecolor{OrangeRed}{rgb}{1.00,0.27,0.00}
\definecolor{PaleGoldenrod}{rgb}{0.93,0.91,0.67}
\definecolor{PaleGreen1}{rgb}{0.60,1.00,0.60}
\definecolor{PaleGreen2}{rgb}{0.56,0.93,0.56}
\definecolor{PaleGreen3}{rgb}{0.49,0.80,0.49}
\definecolor{PaleGreen4}{rgb}{0.33,0.55,0.33}
\definecolor{PaleGreen}{rgb}{0.60,0.98,0.60}
\definecolor{PaleTurquoise1}{rgb}{0.73,1.00,1.00}
\definecolor{PaleTurquoise2}{rgb}{0.68,0.93,0.93}
\definecolor{PaleTurquoise3}{rgb}{0.59,0.80,0.80}
\definecolor{PaleTurquoise4}{rgb}{0.40,0.55,0.55}
\definecolor{PaleTurquoise}{rgb}{0.69,0.93,0.93}
\definecolor{PaleVioletRed1}{rgb}{1.00,0.51,0.67}
\definecolor{PaleVioletRed2}{rgb}{0.93,0.47,0.62}
\definecolor{PaleVioletRed3}{rgb}{0.80,0.41,0.54}
\definecolor{PaleVioletRed4}{rgb}{0.55,0.28,0.36}
\definecolor{PaleVioletRed}{rgb}{0.86,0.44,0.58}
\definecolor{PapayaWhip}{rgb}{1.00,0.94,0.84}
\definecolor{PeachPuff1}{rgb}{1.00,0.85,0.73}
\definecolor{PeachPuff2}{rgb}{0.93,0.80,0.68}
\definecolor{PeachPuff3}{rgb}{0.80,0.69,0.58}
\definecolor{PeachPuff4}{rgb}{0.55,0.47,0.40}
\definecolor{PeachPuff}{rgb}{1.00,0.85,0.73}
\definecolor{PowderBlue}{rgb}{0.69,0.88,0.90}
\definecolor{RosyBrown1}{rgb}{1.00,0.76,0.76}
\definecolor{RosyBrown2}{rgb}{0.93,0.71,0.71}
\definecolor{RosyBrown3}{rgb}{0.80,0.61,0.61}
\definecolor{RosyBrown4}{rgb}{0.55,0.41,0.41}
\definecolor{RosyBrown}{rgb}{0.74,0.56,0.56}
\definecolor{RoyalBlue1}{rgb}{0.28,0.46,1.00}
\definecolor{RoyalBlue2}{rgb}{0.26,0.43,0.93}
\definecolor{RoyalBlue3}{rgb}{0.23,0.37,0.80}
\definecolor{RoyalBlue4}{rgb}{0.15,0.25,0.55}
\definecolor{RoyalBlue}{rgb}{0.25,0.41,0.88}
\definecolor{SaddleBrown}{rgb}{0.55,0.27,0.07}
\definecolor{SandyBrown}{rgb}{0.96,0.64,0.38}
\definecolor{SeaGreen1}{rgb}{0.33,1.00,0.62}
\definecolor{SeaGreen2}{rgb}{0.31,0.93,0.58}
\definecolor{SeaGreen3}{rgb}{0.26,0.80,0.50}
\definecolor{SeaGreen4}{rgb}{0.18,0.55,0.34}
\definecolor{SeaGreen}{rgb}{0.18,0.55,0.34}
\definecolor{SkyBlue1}{rgb}{0.53,0.81,1.00}
\definecolor{SkyBlue2}{rgb}{0.49,0.75,0.93}
\definecolor{SkyBlue3}{rgb}{0.42,0.65,0.80}
\definecolor{SkyBlue4}{rgb}{0.29,0.44,0.55}
\definecolor{SkyBlue}{rgb}{0.53,0.81,0.92}
\definecolor{SlateBlue1}{rgb}{0.51,0.44,1.00}
\definecolor{SlateBlue2}{rgb}{0.48,0.40,0.93}
\definecolor{SlateBlue3}{rgb}{0.41,0.35,0.80}
\definecolor{SlateBlue4}{rgb}{0.28,0.24,0.55}
\definecolor{SlateBlue}{rgb}{0.42,0.35,0.80}
\definecolor{SlateGray1}{rgb}{0.78,0.89,1.00}
\definecolor{SlateGray2}{rgb}{0.73,0.83,0.93}
\definecolor{SlateGray3}{rgb}{0.62,0.71,0.80}
\definecolor{SlateGray4}{rgb}{0.42,0.48,0.55}
\definecolor{SlateGray}{rgb}{0.44,0.50,0.56}
\definecolor{SlateGrey}{rgb}{0.44,0.50,0.56}
\definecolor{SpringGreen1}{rgb}{0.00,1.00,0.50}
\definecolor{SpringGreen2}{rgb}{0.00,0.93,0.46}
\definecolor{SpringGreen3}{rgb}{0.00,0.80,0.40}
\definecolor{SpringGreen4}{rgb}{0.00,0.55,0.27}
\definecolor{SpringGreen}{rgb}{0.00,1.00,0.50}
\definecolor{SteelBlue1}{rgb}{0.39,0.72,1.00}
\definecolor{SteelBlue2}{rgb}{0.36,0.67,0.93}
\definecolor{SteelBlue3}{rgb}{0.31,0.58,0.80}
\definecolor{SteelBlue4}{rgb}{0.21,0.39,0.55}
\definecolor{SteelBlue}{rgb}{0.27,0.51,0.71}
\definecolor{VioletRed1}{rgb}{1.00,0.24,0.59}
\definecolor{VioletRed2}{rgb}{0.93,0.23,0.55}
\definecolor{VioletRed3}{rgb}{0.80,0.20,0.47}
\definecolor{VioletRed4}{rgb}{0.55,0.13,0.32}
\definecolor{VioletRed}{rgb}{0.82,0.13,0.56}
\definecolor{WhiteSmoke}{rgb}{0.96,0.96,0.96}
\definecolor{YellowGreen}{rgb}{0.60,0.80,0.20}
\definecolor{aliceblue}{rgb}{0.94,0.97,1.00}
\definecolor{antiquewhite}{rgb}{0.98,0.92,0.84}
\definecolor{aquamarine1}{rgb}{0.50,1.00,0.83}
\definecolor{aquamarine2}{rgb}{0.46,0.93,0.78}
\definecolor{aquamarine3}{rgb}{0.40,0.80,0.67}
\definecolor{aquamarine4}{rgb}{0.27,0.55,0.45}
\definecolor{aquamarine}{rgb}{0.50,1.00,0.83}
\definecolor{azure1}{rgb}{0.94,1.00,1.00}
\definecolor{azure2}{rgb}{0.88,0.93,0.93}
\definecolor{azure3}{rgb}{0.76,0.80,0.80}
\definecolor{azure4}{rgb}{0.51,0.55,0.55}
\definecolor{azure}{rgb}{0.94,1.00,1.00}
\definecolor{beige}{rgb}{0.96,0.96,0.86}
\definecolor{bisque1}{rgb}{1.00,0.89,0.77}
\definecolor{bisque2}{rgb}{0.93,0.84,0.72}
\definecolor{bisque3}{rgb}{0.80,0.72,0.62}
\definecolor{bisque4}{rgb}{0.55,0.49,0.42}
\definecolor{bisque}{rgb}{1.00,0.89,0.77}
\definecolor{black}{rgb}{0.00,0.00,0.00}
\definecolor{blanchedalmond}{rgb}{1.00,0.92,0.80}
\definecolor{blue1}{rgb}{0.00,0.00,1.00}
\definecolor{blue2}{rgb}{0.00,0.00,0.93}
\definecolor{blue3}{rgb}{0.00,0.00,0.80}
\definecolor{blue4}{rgb}{0.00,0.00,0.55}
\definecolor{blueviolet}{rgb}{0.54,0.17,0.89}
\definecolor{blue}{rgb}{0.00,0.00,1.00}
\definecolor{brown1}{rgb}{1.00,0.25,0.25}
\definecolor{brown2}{rgb}{0.93,0.23,0.23}
\definecolor{brown3}{rgb}{0.80,0.20,0.20}
\definecolor{brown4}{rgb}{0.55,0.14,0.14}
\definecolor{brown}{rgb}{0.65,0.16,0.16}
\definecolor{burlywood1}{rgb}{1.00,0.83,0.61}
\definecolor{burlywood2}{rgb}{0.93,0.77,0.57}
\definecolor{burlywood3}{rgb}{0.80,0.67,0.49}
\definecolor{burlywood4}{rgb}{0.55,0.45,0.33}
\definecolor{burlywood}{rgb}{0.87,0.72,0.53}
\definecolor{cadetblue}{rgb}{0.37,0.62,0.63}
\definecolor{chartreuse1}{rgb}{0.50,1.00,0.00}
\definecolor{chartreuse2}{rgb}{0.46,0.93,0.00}
\definecolor{chartreuse3}{rgb}{0.40,0.80,0.00}
\definecolor{chartreuse4}{rgb}{0.27,0.55,0.00}
\definecolor{chartreuse}{rgb}{0.50,1.00,0.00}
\definecolor{chocolate1}{rgb}{1.00,0.50,0.14}
\definecolor{chocolate2}{rgb}{0.93,0.46,0.13}
\definecolor{chocolate3}{rgb}{0.80,0.40,0.11}
\definecolor{chocolate4}{rgb}{0.55,0.27,0.07}
\definecolor{chocolate}{rgb}{0.82,0.41,0.12}
\definecolor{coral1}{rgb}{1.00,0.45,0.34}
\definecolor{coral2}{rgb}{0.93,0.42,0.31}
\definecolor{coral3}{rgb}{0.80,0.36,0.27}
\definecolor{coral4}{rgb}{0.55,0.24,0.18}
\definecolor{coral}{rgb}{1.00,0.50,0.31}
\definecolor{cornflowerblue}{rgb}{0.39,0.58,0.93}
\definecolor{cornsilk1}{rgb}{1.00,0.97,0.86}
\definecolor{cornsilk2}{rgb}{0.93,0.91,0.80}
\definecolor{cornsilk3}{rgb}{0.80,0.78,0.69}
\definecolor{cornsilk4}{rgb}{0.55,0.53,0.47}
\definecolor{cornsilk}{rgb}{1.00,0.97,0.86}
\definecolor{cyan1}{rgb}{0.00,1.00,1.00}
\definecolor{cyan2}{rgb}{0.00,0.93,0.93}
\definecolor{cyan3}{rgb}{0.00,0.80,0.80}
\definecolor{cyan4}{rgb}{0.00,0.55,0.55}
\definecolor{cyan}{rgb}{0.00,1.00,1.00}
\definecolor{darkblue}{rgb}{0.00,0.00,0.55}
\definecolor{darkcyan}{rgb}{0.00,0.55,0.55}
\definecolor{darkgoldenrod}{rgb}{0.72,0.53,0.04}
\definecolor{darkgray}{rgb}{0.66,0.66,0.66}
\definecolor{darkgreen}{rgb}{0.00,0.39,0.00}
\definecolor{darkgrey}{rgb}{0.66,0.66,0.66}
\definecolor{darkkhaki}{rgb}{0.74,0.72,0.42}
\definecolor{darkmagenta}{rgb}{0.55,0.00,0.55}
\definecolor{darkolive}{rgb}{0.33,0.42,0.18}
\definecolor{darkorange}{rgb}{1.00,0.55,0.00}
\definecolor{darkorchid}{rgb}{0.60,0.20,0.80}
\definecolor{darkred}{rgb}{0.55,0.00,0.00}
\definecolor{darksalmon}{rgb}{0.91,0.59,0.48}
\definecolor{darksea}{rgb}{0.56,0.74,0.56}
\definecolor{darkslate}{rgb}{0.18,0.31,0.31}
\definecolor{darkslate}{rgb}{0.18,0.31,0.31}
\definecolor{darkslate}{rgb}{0.28,0.24,0.55}
\definecolor{darkturquoise}{rgb}{0.00,0.81,0.82}
\definecolor{darkviolet}{rgb}{0.58,0.00,0.83}
\definecolor{deeppink}{rgb}{1.00,0.08,0.58}
\definecolor{deepsky}{rgb}{0.00,0.75,1.00}
\definecolor{dimgray}{rgb}{0.41,0.41,0.41}
\definecolor{dimgrey}{rgb}{0.41,0.41,0.41}
\definecolor{dodgerblue}{rgb}{0.12,0.56,1.00}
\definecolor{firebrick1}{rgb}{1.00,0.19,0.19}
\definecolor{firebrick2}{rgb}{0.93,0.17,0.17}
\definecolor{firebrick3}{rgb}{0.80,0.15,0.15}
\definecolor{firebrick4}{rgb}{0.55,0.10,0.10}
\definecolor{firebrick}{rgb}{0.70,0.13,0.13}
\definecolor{floralwhite}{rgb}{1.00,0.98,0.94}
\definecolor{forestgreen}{rgb}{0.13,0.55,0.13}
\definecolor{gainsboro}{rgb}{0.86,0.86,0.86}
\definecolor{ghostwhite}{rgb}{0.97,0.97,1.00}
\definecolor{gold1}{rgb}{1.00,0.84,0.00}
\definecolor{gold2}{rgb}{0.93,0.79,0.00}
\definecolor{gold3}{rgb}{0.80,0.68,0.00}
\definecolor{gold4}{rgb}{0.55,0.46,0.00}
\definecolor{goldenrod1}{rgb}{1.00,0.76,0.15}
\definecolor{goldenrod2}{rgb}{0.93,0.71,0.13}
\definecolor{goldenrod3}{rgb}{0.80,0.61,0.11}
\definecolor{goldenrod4}{rgb}{0.55,0.41,0.08}
\definecolor{goldenrod}{rgb}{0.85,0.65,0.13}
\definecolor{gold}{rgb}{1.00,0.84,0.00}
\definecolor{gray0}{rgb}{0.00,0.00,0.00}
\definecolor{gray100}{rgb}{1.00,1.00,1.00}
\definecolor{gray10}{rgb}{0.10,0.10,0.10}
\definecolor{gray11}{rgb}{0.11,0.11,0.11}
\definecolor{gray12}{rgb}{0.12,0.12,0.12}
\definecolor{gray13}{rgb}{0.13,0.13,0.13}
\definecolor{gray14}{rgb}{0.14,0.14,0.14}
\definecolor{gray15}{rgb}{0.15,0.15,0.15}
\definecolor{gray16}{rgb}{0.16,0.16,0.16}
\definecolor{gray17}{rgb}{0.17,0.17,0.17}
\definecolor{gray18}{rgb}{0.18,0.18,0.18}
\definecolor{gray19}{rgb}{0.19,0.19,0.19}
\definecolor{gray1}{rgb}{0.01,0.01,0.01}
\definecolor{gray20}{rgb}{0.20,0.20,0.20}
\definecolor{gray21}{rgb}{0.21,0.21,0.21}
\definecolor{gray22}{rgb}{0.22,0.22,0.22}
\definecolor{gray23}{rgb}{0.23,0.23,0.23}
\definecolor{gray24}{rgb}{0.24,0.24,0.24}
\definecolor{gray25}{rgb}{0.25,0.25,0.25}
\definecolor{gray26}{rgb}{0.26,0.26,0.26}
\definecolor{gray27}{rgb}{0.27,0.27,0.27}
\definecolor{gray28}{rgb}{0.28,0.28,0.28}
\definecolor{gray29}{rgb}{0.29,0.29,0.29}
\definecolor{gray2}{rgb}{0.02,0.02,0.02}
\definecolor{gray30}{rgb}{0.30,0.30,0.30}
\definecolor{gray31}{rgb}{0.31,0.31,0.31}
\definecolor{gray32}{rgb}{0.32,0.32,0.32}
\definecolor{gray33}{rgb}{0.33,0.33,0.33}
\definecolor{gray34}{rgb}{0.34,0.34,0.34}
\definecolor{gray35}{rgb}{0.35,0.35,0.35}
\definecolor{gray36}{rgb}{0.36,0.36,0.36}
\definecolor{gray37}{rgb}{0.37,0.37,0.37}
\definecolor{gray38}{rgb}{0.38,0.38,0.38}
\definecolor{gray39}{rgb}{0.39,0.39,0.39}
\definecolor{gray3}{rgb}{0.03,0.03,0.03}
\definecolor{gray40}{rgb}{0.40,0.40,0.40}
\definecolor{gray41}{rgb}{0.41,0.41,0.41}
\definecolor{gray42}{rgb}{0.42,0.42,0.42}
\definecolor{gray43}{rgb}{0.43,0.43,0.43}
\definecolor{gray44}{rgb}{0.44,0.44,0.44}
\definecolor{gray45}{rgb}{0.45,0.45,0.45}
\definecolor{gray46}{rgb}{0.46,0.46,0.46}
\definecolor{gray47}{rgb}{0.47,0.47,0.47}
\definecolor{gray48}{rgb}{0.48,0.48,0.48}
\definecolor{gray49}{rgb}{0.49,0.49,0.49}
\definecolor{gray4}{rgb}{0.04,0.04,0.04}
\definecolor{gray50}{rgb}{0.50,0.50,0.50}
\definecolor{gray51}{rgb}{0.51,0.51,0.51}
\definecolor{gray52}{rgb}{0.52,0.52,0.52}
\definecolor{gray53}{rgb}{0.53,0.53,0.53}
\definecolor{gray54}{rgb}{0.54,0.54,0.54}
\definecolor{gray55}{rgb}{0.55,0.55,0.55}
\definecolor{gray56}{rgb}{0.56,0.56,0.56}
\definecolor{gray57}{rgb}{0.57,0.57,0.57}
\definecolor{gray58}{rgb}{0.58,0.58,0.58}
\definecolor{gray59}{rgb}{0.59,0.59,0.59}
\definecolor{gray5}{rgb}{0.05,0.05,0.05}
\definecolor{gray60}{rgb}{0.60,0.60,0.60}
\definecolor{gray61}{rgb}{0.61,0.61,0.61}
\definecolor{gray62}{rgb}{0.62,0.62,0.62}
\definecolor{gray63}{rgb}{0.63,0.63,0.63}
\definecolor{gray64}{rgb}{0.64,0.64,0.64}
\definecolor{gray65}{rgb}{0.65,0.65,0.65}
\definecolor{gray66}{rgb}{0.66,0.66,0.66}
\definecolor{gray67}{rgb}{0.67,0.67,0.67}
\definecolor{gray68}{rgb}{0.68,0.68,0.68}
\definecolor{gray69}{rgb}{0.69,0.69,0.69}
\definecolor{gray6}{rgb}{0.06,0.06,0.06}
\definecolor{gray70}{rgb}{0.70,0.70,0.70}
\definecolor{gray71}{rgb}{0.71,0.71,0.71}
\definecolor{gray72}{rgb}{0.72,0.72,0.72}
\definecolor{gray73}{rgb}{0.73,0.73,0.73}
\definecolor{gray74}{rgb}{0.74,0.74,0.74}
\definecolor{gray75}{rgb}{0.75,0.75,0.75}
\definecolor{gray76}{rgb}{0.76,0.76,0.76}
\definecolor{gray77}{rgb}{0.77,0.77,0.77}
\definecolor{gray78}{rgb}{0.78,0.78,0.78}
\definecolor{gray79}{rgb}{0.79,0.79,0.79}
\definecolor{gray7}{rgb}{0.07,0.07,0.07}
\definecolor{gray80}{rgb}{0.80,0.80,0.80}
\definecolor{gray81}{rgb}{0.81,0.81,0.81}
\definecolor{gray82}{rgb}{0.82,0.82,0.82}
\definecolor{gray83}{rgb}{0.83,0.83,0.83}
\definecolor{gray84}{rgb}{0.84,0.84,0.84}
\definecolor{gray85}{rgb}{0.85,0.85,0.85}
\definecolor{gray86}{rgb}{0.86,0.86,0.86}
\definecolor{gray87}{rgb}{0.87,0.87,0.87}
\definecolor{gray88}{rgb}{0.88,0.88,0.88}
\definecolor{gray89}{rgb}{0.89,0.89,0.89}
\definecolor{gray8}{rgb}{0.08,0.08,0.08}
\definecolor{gray90}{rgb}{0.90,0.90,0.90}
\definecolor{gray91}{rgb}{0.91,0.91,0.91}
\definecolor{gray92}{rgb}{0.92,0.92,0.92}
\definecolor{gray93}{rgb}{0.93,0.93,0.93}
\definecolor{gray94}{rgb}{0.94,0.94,0.94}
\definecolor{gray95}{rgb}{0.95,0.95,0.95}
\definecolor{gray96}{rgb}{0.96,0.96,0.96}
\definecolor{gray97}{rgb}{0.97,0.97,0.97}
\definecolor{gray98}{rgb}{0.98,0.98,0.98}
\definecolor{gray99}{rgb}{0.99,0.99,0.99}
\definecolor{gray9}{rgb}{0.09,0.09,0.09}
\definecolor{gray}{rgb}{0.75,0.75,0.75}
\definecolor{green1}{rgb}{0.00,1.00,0.00}
\definecolor{green2}{rgb}{0.00,0.93,0.00}
\definecolor{green3}{rgb}{0.00,0.80,0.00}
\definecolor{green4}{rgb}{0.00,0.55,0.00}
\definecolor{greenyellow}{rgb}{0.68,1.00,0.18}
\definecolor{green}{rgb}{0.00,1.00,0.00}
\definecolor{grey0}{rgb}{0.00,0.00,0.00}
\definecolor{grey100}{rgb}{1.00,1.00,1.00}
\definecolor{grey10}{rgb}{0.10,0.10,0.10}
\definecolor{grey11}{rgb}{0.11,0.11,0.11}
\definecolor{grey12}{rgb}{0.12,0.12,0.12}
\definecolor{grey13}{rgb}{0.13,0.13,0.13}
\definecolor{grey14}{rgb}{0.14,0.14,0.14}
\definecolor{grey15}{rgb}{0.15,0.15,0.15}
\definecolor{grey16}{rgb}{0.16,0.16,0.16}
\definecolor{grey17}{rgb}{0.17,0.17,0.17}
\definecolor{grey18}{rgb}{0.18,0.18,0.18}
\definecolor{grey19}{rgb}{0.19,0.19,0.19}
\definecolor{grey1}{rgb}{0.01,0.01,0.01}
\definecolor{grey20}{rgb}{0.20,0.20,0.20}
\definecolor{grey21}{rgb}{0.21,0.21,0.21}
\definecolor{grey22}{rgb}{0.22,0.22,0.22}
\definecolor{grey23}{rgb}{0.23,0.23,0.23}
\definecolor{grey24}{rgb}{0.24,0.24,0.24}
\definecolor{grey25}{rgb}{0.25,0.25,0.25}
\definecolor{grey26}{rgb}{0.26,0.26,0.26}
\definecolor{grey27}{rgb}{0.27,0.27,0.27}
\definecolor{grey28}{rgb}{0.28,0.28,0.28}
\definecolor{grey29}{rgb}{0.29,0.29,0.29}
\definecolor{grey2}{rgb}{0.02,0.02,0.02}
\definecolor{grey30}{rgb}{0.30,0.30,0.30}
\definecolor{grey31}{rgb}{0.31,0.31,0.31}
\definecolor{grey32}{rgb}{0.32,0.32,0.32}
\definecolor{grey33}{rgb}{0.33,0.33,0.33}
\definecolor{grey34}{rgb}{0.34,0.34,0.34}
\definecolor{grey35}{rgb}{0.35,0.35,0.35}
\definecolor{grey36}{rgb}{0.36,0.36,0.36}
\definecolor{grey37}{rgb}{0.37,0.37,0.37}
\definecolor{grey38}{rgb}{0.38,0.38,0.38}
\definecolor{grey39}{rgb}{0.39,0.39,0.39}
\definecolor{grey3}{rgb}{0.03,0.03,0.03}
\definecolor{grey40}{rgb}{0.40,0.40,0.40}
\definecolor{grey41}{rgb}{0.41,0.41,0.41}
\definecolor{grey42}{rgb}{0.42,0.42,0.42}
\definecolor{grey43}{rgb}{0.43,0.43,0.43}
\definecolor{grey44}{rgb}{0.44,0.44,0.44}
\definecolor{grey45}{rgb}{0.45,0.45,0.45}
\definecolor{grey46}{rgb}{0.46,0.46,0.46}
\definecolor{grey47}{rgb}{0.47,0.47,0.47}
\definecolor{grey48}{rgb}{0.48,0.48,0.48}
\definecolor{grey49}{rgb}{0.49,0.49,0.49}
\definecolor{grey4}{rgb}{0.04,0.04,0.04}
\definecolor{grey50}{rgb}{0.50,0.50,0.50}
\definecolor{grey51}{rgb}{0.51,0.51,0.51}
\definecolor{grey52}{rgb}{0.52,0.52,0.52}
\definecolor{grey53}{rgb}{0.53,0.53,0.53}
\definecolor{grey54}{rgb}{0.54,0.54,0.54}
\definecolor{grey55}{rgb}{0.55,0.55,0.55}
\definecolor{grey56}{rgb}{0.56,0.56,0.56}
\definecolor{grey57}{rgb}{0.57,0.57,0.57}
\definecolor{grey58}{rgb}{0.58,0.58,0.58}
\definecolor{grey59}{rgb}{0.59,0.59,0.59}
\definecolor{grey5}{rgb}{0.05,0.05,0.05}
\definecolor{grey60}{rgb}{0.60,0.60,0.60}
\definecolor{grey61}{rgb}{0.61,0.61,0.61}
\definecolor{grey62}{rgb}{0.62,0.62,0.62}
\definecolor{grey63}{rgb}{0.63,0.63,0.63}
\definecolor{grey64}{rgb}{0.64,0.64,0.64}
\definecolor{grey65}{rgb}{0.65,0.65,0.65}
\definecolor{grey66}{rgb}{0.66,0.66,0.66}
\definecolor{grey67}{rgb}{0.67,0.67,0.67}
\definecolor{grey68}{rgb}{0.68,0.68,0.68}
\definecolor{grey69}{rgb}{0.69,0.69,0.69}
\definecolor{grey6}{rgb}{0.06,0.06,0.06}
\definecolor{grey70}{rgb}{0.70,0.70,0.70}
\definecolor{grey71}{rgb}{0.71,0.71,0.71}
\definecolor{grey72}{rgb}{0.72,0.72,0.72}
\definecolor{grey73}{rgb}{0.73,0.73,0.73}
\definecolor{grey74}{rgb}{0.74,0.74,0.74}
\definecolor{grey75}{rgb}{0.75,0.75,0.75}
\definecolor{grey76}{rgb}{0.76,0.76,0.76}
\definecolor{grey77}{rgb}{0.77,0.77,0.77}
\definecolor{grey78}{rgb}{0.78,0.78,0.78}
\definecolor{grey79}{rgb}{0.79,0.79,0.79}
\definecolor{grey7}{rgb}{0.07,0.07,0.07}
\definecolor{grey80}{rgb}{0.80,0.80,0.80}
\definecolor{grey81}{rgb}{0.81,0.81,0.81}
\definecolor{grey82}{rgb}{0.82,0.82,0.82}
\definecolor{grey83}{rgb}{0.83,0.83,0.83}
\definecolor{grey84}{rgb}{0.84,0.84,0.84}
\definecolor{grey85}{rgb}{0.85,0.85,0.85}
\definecolor{grey86}{rgb}{0.86,0.86,0.86}
\definecolor{grey87}{rgb}{0.87,0.87,0.87}
\definecolor{grey88}{rgb}{0.88,0.88,0.88}
\definecolor{grey89}{rgb}{0.89,0.89,0.89}
\definecolor{grey8}{rgb}{0.08,0.08,0.08}
\definecolor{grey90}{rgb}{0.90,0.90,0.90}
\definecolor{grey91}{rgb}{0.91,0.91,0.91}
\definecolor{grey92}{rgb}{0.92,0.92,0.92}
\definecolor{grey93}{rgb}{0.93,0.93,0.93}
\definecolor{grey94}{rgb}{0.94,0.94,0.94}
\definecolor{grey95}{rgb}{0.95,0.95,0.95}
\definecolor{grey96}{rgb}{0.96,0.96,0.96}
\definecolor{grey97}{rgb}{0.97,0.97,0.97}
\definecolor{grey98}{rgb}{0.98,0.98,0.98}
\definecolor{grey99}{rgb}{0.99,0.99,0.99}
\definecolor{grey9}{rgb}{0.09,0.09,0.09}
\definecolor{grey}{rgb}{0.75,0.75,0.75}
\definecolor{honeydew1}{rgb}{0.94,1.00,0.94}
\definecolor{honeydew2}{rgb}{0.88,0.93,0.88}
\definecolor{honeydew3}{rgb}{0.76,0.80,0.76}
\definecolor{honeydew4}{rgb}{0.51,0.55,0.51}
\definecolor{honeydew}{rgb}{0.94,1.00,0.94}
\definecolor{hotpink}{rgb}{1.00,0.41,0.71}
\definecolor{indianred}{rgb}{0.80,0.36,0.36}
\definecolor{ivory1}{rgb}{1.00,1.00,0.94}
\definecolor{ivory2}{rgb}{0.93,0.93,0.88}
\definecolor{ivory3}{rgb}{0.80,0.80,0.76}
\definecolor{ivory4}{rgb}{0.55,0.55,0.51}
\definecolor{ivory}{rgb}{1.00,1.00,0.94}
\definecolor{khaki1}{rgb}{1.00,0.96,0.56}
\definecolor{khaki2}{rgb}{0.93,0.90,0.52}
\definecolor{khaki3}{rgb}{0.80,0.78,0.45}
\definecolor{khaki4}{rgb}{0.55,0.53,0.31}
\definecolor{khaki}{rgb}{0.94,0.90,0.55}
\definecolor{lavenderblush}{rgb}{1.00,0.94,0.96}
\definecolor{lavender}{rgb}{0.90,0.90,0.98}
\definecolor{lawngreen}{rgb}{0.49,0.99,0.00}
\definecolor{lemonchiffon}{rgb}{1.00,0.98,0.80}
\definecolor{lightblue}{rgb}{0.68,0.85,0.90}
\definecolor{lightcoral}{rgb}{0.94,0.50,0.50}
\definecolor{lightcyan}{rgb}{0.88,1.00,1.00}
\definecolor{lightgoldenrod}{rgb}{0.93,0.87,0.51}
\definecolor{lightgoldenrod}{rgb}{0.98,0.98,0.82}
\definecolor{lightgray}{rgb}{0.83,0.83,0.83}
\definecolor{lightgreen}{rgb}{0.56,0.93,0.56}
\definecolor{lightgrey}{rgb}{0.83,0.83,0.83}
\definecolor{lightpink}{rgb}{1.00,0.71,0.76}
\definecolor{lightsalmon}{rgb}{1.00,0.63,0.48}
\definecolor{lightsea}{rgb}{0.13,0.70,0.67}
\definecolor{lightsky}{rgb}{0.53,0.81,0.98}
\definecolor{lightslate}{rgb}{0.47,0.53,0.60}
\definecolor{lightslate}{rgb}{0.47,0.53,0.60}
\definecolor{lightslate}{rgb}{0.52,0.44,1.00}
\definecolor{lightsteel}{rgb}{0.69,0.77,0.87}
\definecolor{lightyellow}{rgb}{1.00,1.00,0.88}
\definecolor{limegreen}{rgb}{0.20,0.80,0.20}
\definecolor{linen}{rgb}{0.98,0.94,0.90}
\definecolor{magenta1}{rgb}{1.00,0.00,1.00}
\definecolor{magenta2}{rgb}{0.93,0.00,0.93}
\definecolor{magenta3}{rgb}{0.80,0.00,0.80}
\definecolor{magenta4}{rgb}{0.55,0.00,0.55}
\definecolor{magenta}{rgb}{1.00,0.00,1.00}
\definecolor{maroon1}{rgb}{1.00,0.20,0.70}
\definecolor{maroon2}{rgb}{0.93,0.19,0.65}
\definecolor{maroon3}{rgb}{0.80,0.16,0.56}
\definecolor{maroon4}{rgb}{0.55,0.11,0.38}
\definecolor{maroon}{rgb}{0.69,0.19,0.38}
\definecolor{mediumaquamarine}{rgb}{0.40,0.80,0.67}
\definecolor{mediumblue}{rgb}{0.00,0.00,0.80}
\definecolor{mediumorchid}{rgb}{0.73,0.33,0.83}
\definecolor{mediumpurple}{rgb}{0.58,0.44,0.86}
\definecolor{mediumsea}{rgb}{0.24,0.70,0.44}
\definecolor{mediumslate}{rgb}{0.48,0.41,0.93}
\definecolor{mediumspring}{rgb}{0.00,0.98,0.60}
\definecolor{mediumturquoise}{rgb}{0.28,0.82,0.80}
\definecolor{mediumviolet}{rgb}{0.78,0.08,0.52}
\definecolor{midnightblue}{rgb}{0.10,0.10,0.44}
\definecolor{mintcream}{rgb}{0.96,1.00,0.98}
\definecolor{mistyrose}{rgb}{1.00,0.89,0.88}
\definecolor{moccasin}{rgb}{1.00,0.89,0.71}
\definecolor{navajowhite}{rgb}{1.00,0.87,0.68}
\definecolor{navyblue}{rgb}{0.00,0.00,0.50}
\definecolor{navy}{rgb}{0.00,0.00,0.50}
\definecolor{oldlace}{rgb}{0.99,0.96,0.90}
\definecolor{olivedrab}{rgb}{0.42,0.56,0.14}
\definecolor{orange1}{rgb}{1.00,0.65,0.00}
\definecolor{orange2}{rgb}{0.93,0.60,0.00}
\definecolor{orange3}{rgb}{0.80,0.52,0.00}
\definecolor{orange4}{rgb}{0.55,0.35,0.00}
\definecolor{orangered}{rgb}{1.00,0.27,0.00}
\definecolor{orange}{rgb}{1.00,0.65,0.00}
\definecolor{orchid1}{rgb}{1.00,0.51,0.98}
\definecolor{orchid2}{rgb}{0.93,0.48,0.91}
\definecolor{orchid3}{rgb}{0.80,0.41,0.79}
\definecolor{orchid4}{rgb}{0.55,0.28,0.54}
\definecolor{orchid}{rgb}{0.85,0.44,0.84}
\definecolor{palegoldenrod}{rgb}{0.93,0.91,0.67}
\definecolor{palegreen}{rgb}{0.60,0.98,0.60}
\definecolor{paleturquoise}{rgb}{0.69,0.93,0.93}
\definecolor{paleviolet}{rgb}{0.86,0.44,0.58}
\definecolor{papayawhip}{rgb}{1.00,0.94,0.84}
\definecolor{peachpuff}{rgb}{1.00,0.85,0.73}
\definecolor{peru}{rgb}{0.80,0.52,0.25}
\definecolor{pink1}{rgb}{1.00,0.71,0.77}
\definecolor{pink2}{rgb}{0.93,0.66,0.72}
\definecolor{pink3}{rgb}{0.80,0.57,0.62}
\definecolor{pink4}{rgb}{0.55,0.39,0.42}
\definecolor{pink}{rgb}{1.00,0.75,0.80}
\definecolor{plum1}{rgb}{1.00,0.73,1.00}
\definecolor{plum2}{rgb}{0.93,0.68,0.93}
\definecolor{plum3}{rgb}{0.80,0.59,0.80}
\definecolor{plum4}{rgb}{0.55,0.40,0.55}
\definecolor{plum}{rgb}{0.87,0.63,0.87}
\definecolor{powderblue}{rgb}{0.69,0.88,0.90}
\definecolor{purple1}{rgb}{0.61,0.19,1.00}
\definecolor{purple2}{rgb}{0.57,0.17,0.93}
\definecolor{purple3}{rgb}{0.49,0.15,0.80}
\definecolor{purple4}{rgb}{0.33,0.10,0.55}
\definecolor{purple}{rgb}{0.63,0.13,0.94}
\definecolor{red1}{rgb}{1.00,0.00,0.00}
\definecolor{red2}{rgb}{0.93,0.00,0.00}
\definecolor{red3}{rgb}{0.80,0.00,0.00}
\definecolor{red4}{rgb}{0.55,0.00,0.00}
\definecolor{red}{rgb}{1.00,0.00,0.00}
\definecolor{rosybrown}{rgb}{0.74,0.56,0.56}
\definecolor{royalblue}{rgb}{0.25,0.41,0.88}
\definecolor{saddlebrown}{rgb}{0.55,0.27,0.07}
\definecolor{salmon1}{rgb}{1.00,0.55,0.41}
\definecolor{salmon2}{rgb}{0.93,0.51,0.38}
\definecolor{salmon3}{rgb}{0.80,0.44,0.33}
\definecolor{salmon4}{rgb}{0.55,0.30,0.22}
\definecolor{salmon}{rgb}{0.98,0.50,0.45}
\definecolor{sandybrown}{rgb}{0.96,0.64,0.38}
\definecolor{seagreen}{rgb}{0.18,0.55,0.34}
\definecolor{seashell1}{rgb}{1.00,0.96,0.93}
\definecolor{seashell2}{rgb}{0.93,0.90,0.87}
\definecolor{seashell3}{rgb}{0.80,0.77,0.75}
\definecolor{seashell4}{rgb}{0.55,0.53,0.51}
\definecolor{seashell}{rgb}{1.00,0.96,0.93}
\definecolor{sienna1}{rgb}{1.00,0.51,0.28}
\definecolor{sienna2}{rgb}{0.93,0.47,0.26}
\definecolor{sienna3}{rgb}{0.80,0.41,0.22}
\definecolor{sienna4}{rgb}{0.55,0.28,0.15}
\definecolor{sienna}{rgb}{0.63,0.32,0.18}
\definecolor{skyblue}{rgb}{0.53,0.81,0.92}
\definecolor{slateblue}{rgb}{0.42,0.35,0.80}
\definecolor{slategray}{rgb}{0.44,0.50,0.56}
\definecolor{slategrey}{rgb}{0.44,0.50,0.56}
\definecolor{snow1}{rgb}{1.00,0.98,0.98}
\definecolor{snow2}{rgb}{0.93,0.91,0.91}
\definecolor{snow3}{rgb}{0.80,0.79,0.79}
\definecolor{snow4}{rgb}{0.55,0.54,0.54}
\definecolor{snow}{rgb}{1.00,0.98,0.98}
\definecolor{springgreen}{rgb}{0.00,1.00,0.50}
\definecolor{steelblue}{rgb}{0.27,0.51,0.71}
\definecolor{tan1}{rgb}{1.00,0.65,0.31}
\definecolor{tan2}{rgb}{0.93,0.60,0.29}
\definecolor{tan3}{rgb}{0.80,0.52,0.25}
\definecolor{tan4}{rgb}{0.55,0.35,0.17}
\definecolor{tan}{rgb}{0.82,0.71,0.55}
\definecolor{thistle1}{rgb}{1.00,0.88,1.00}
\definecolor{thistle2}{rgb}{0.93,0.82,0.93}
\definecolor{thistle3}{rgb}{0.80,0.71,0.80}
\definecolor{thistle4}{rgb}{0.55,0.48,0.55}
\definecolor{thistle}{rgb}{0.85,0.75,0.85}
\definecolor{tomato1}{rgb}{1.00,0.39,0.28}
\definecolor{tomato2}{rgb}{0.93,0.36,0.26}
\definecolor{tomato3}{rgb}{0.80,0.31,0.22}
\definecolor{tomato4}{rgb}{0.55,0.21,0.15}
\definecolor{tomato}{rgb}{1.00,0.39,0.28}
\definecolor{turquoise1}{rgb}{0.00,0.96,1.00}
\definecolor{turquoise2}{rgb}{0.00,0.90,0.93}
\definecolor{turquoise3}{rgb}{0.00,0.77,0.80}
\definecolor{turquoise4}{rgb}{0.00,0.53,0.55}
\definecolor{turquoise}{rgb}{0.25,0.88,0.82}
\definecolor{violetred}{rgb}{0.82,0.13,0.56}
\definecolor{violet}{rgb}{0.93,0.51,0.93}
\definecolor{wheat1}{rgb}{1.00,0.91,0.73}
\definecolor{wheat2}{rgb}{0.93,0.85,0.68}
\definecolor{wheat3}{rgb}{0.80,0.73,0.59}
\definecolor{wheat4}{rgb}{0.55,0.49,0.40}
\definecolor{wheat}{rgb}{0.96,0.87,0.70}
\definecolor{whitesmoke}{rgb}{0.96,0.96,0.96}
\definecolor{white}{rgb}{1.00,1.00,1.00}
\definecolor{yellow1}{rgb}{1.00,1.00,0.00}
\definecolor{yellow2}{rgb}{0.93,0.93,0.00}
\definecolor{yellow3}{rgb}{0.80,0.80,0.00}
\definecolor{yellow4}{rgb}{0.55,0.55,0.00}
\definecolor{yellowgreen}{rgb}{0.60,0.80,0.20}
\definecolor{yellow}{rgb}{1.00,1.00,0.00}

\definecolor{white1}{rgb}{1.00,1.00,1.00}

\definecolor{frontal1}{rgb}{0.93,0.27,0.21}
\definecolor{lateral1}{rgb}{0.68,0.66,0.80}
\definecolor{horizontal1}{rgb}{0.94,0.63,0.37}

\definecolor{tnode1}{rgb}{0.93,0.27,0.21}
\definecolor{mnode1}{rgb}{0.68,0.66,0.80}

\definecolor{unsure1}{rgb}{1.00,0.00,0.46}

\hypersetup{
    pdfpagelayout={},
    pdftitle={Tensor Ensemble Learning for Multidimensional Data},
    pdfsubject={},
    pdfauthor={Ilia Kisil, Ahmad Moniri, Danilo P. Mandic},
    pdfkeywords={Tensor Decomposition,  Multidimensional Data, Ensemble Learning, Classification, Bagging},
    pdfstartview=FitH,
    pdfpagemode={UseOutlines},
    bookmarksnumbered=true, bookmarksopen=true, colorlinks,
    citecolor=black,%
    filecolor=black,%
    linkcolor=black,%
    urlcolor=black
}
\usetikzlibrary{shapes.geometric,
                arrows, arrows.meta,
                calc
}

\title{Tensor Ensemble Learning for Multidimensional Data}

\author[1]{Ilia Kisil}
\author[1]{Ahmad Moniri}
\author[1]{Danilo P. Mandic}
\affil[1]{\small
    Electrical and Electronic Engineering Department, Imperial College London, SW7 2AZ, UK\\
    E-mails: \{i.kisil15, ahmad.moniri13, d.mandic\}@imperial.ac.uk
}

\date{}

\begin{document}
\maketitle

\begin{abstract}
    In big data applications, classical ensemble learning is typically infeasible on the raw input data and dimensionality reduction techniques are necessary.
    To this end, novel framework that generalises classic flat-view ensemble learning to multidimensional tensor-valued data is introduced.
    This is achieved by virtue of tensor decompositions, whereby the proposed method, referred to as tensor ensemble learning (TEL),
    decomposes every input data sample into multiple factors which allows for a flexibility in the choice of multiple learning algorithms
    in order to improve test performance.
    The TEL framework is shown to naturally compress multidimensional data in order to
    take advantage of the inherent multi-way data structure and exploit the benefit of ensemble learning.
    The proposed framework is verified through the application of Higher Order Singular Value Decomposition (HOSVD) to the ETH-80 dataset
    and is shown to outperform the classical ensemble learning approach of bootstrap aggregating.
\end{abstract}

\providecommand{\keywords}[1]{\textbf{\textit{Index terms---}} #1}

\section{Introduction}
    The phenomenon of the \textit{wisdom of the crowd} has been known for a very long time and
    was originally formulated by Aristotle.
    It simply states that the collective answer of a group of people to questions related to
    common world knowledge, spatial reasoning, and general estimation tasks, is often superior to
    the judgement of a particular person within this group.

    With the advent of computer, the machine learning community have adopted this concept under the framework of \textit{ensemble learning} \cite{dietterich2000ensemble}.
    This class of methods can be described as a collection of base learners whereby each learner generates a particular
    hypothesis about an underlying process that governs the input data.
    This makes it possible to construct the final model as a strategic aggregation of the outputs from its constituent learners.
    Indeed, the ``wisdom of the base learners'' has been proven to be a powerful way to enhance the performance
    when solving both classification and regression types of problems \cite{dietterich2000experimental, opitz1999popular, dvzeroski2004combining}.
    For example, ensemble learning has been a key element in winning solutions to the Netflix Prize competition \cite{koren2009bellkor}\notshow{\cite{toscher2009bigchaos}}.
    Generally speaking, ensemble methods can be considered within the three main groups:

    \begin{itemize}[itemindent=0.3cm, leftmargin=0cm]
        \item \textit{Boosting}.
        The aim is to sequentially train a series of estimators whereby every subsequent estimator puts more emphasis
        on samples that were previously predicted incorrectly \cite{schapire1990strength}.
        The most established variant of this strategy is the \textit{AdaBoost} algorithm\cite{freund1997decision}.
        \item \textit{Stacking}.
        The gist of his strategy is that a set of base classifiers is first trained
        on the original input data, followed by ``meta learning'' in order to combine the outputs
        of base classifiers for enhanced performance \cite{wolpert1992stacked}.
        \item \textit{Bagging}. This family relies on creation of multiple ``surrogate'' training datasets from the original data,
        for example through resampling with replacement \cite{breiman1996bagging}.
        An independent base classifier is then applied to each new dataset and their outputs are combined to reach a consensus.
        To date, the most successful bagging approach is the Random Forest algorithm \cite{liaw2002classification}.
    \end{itemize}

    It is important to notice that, common machine learning algorithms, such as Neural Networks (NN) or Support Vector Machines (SVM),
    operate on ``one way'' vector inputs, even if the original data is inherently multidimensional or multivariate, that is ``multi-way''.
    Such ``flat view'' representation of multidimensional data arrays (also called tensors) prevents learning models from taking full
    advantage of the underlying multi-way latent structure, as in the ``matrix world'' the cross-modal dependencies become obscure or even completely broken.
    This all calls for modern methodologies which can maintain the original structure in the
    data, an ideal avenue to explore multi-way analyses based on tensor decompositions \cite{cichocki2015tensor}.
    Indeed, tensor based representations of multi-faceted data have been shown to be a promising tool for feature extraction \cite{wong2015joint, kisil2018common},
    data fusion \cite{acar2013understanding, calvi2018sum},
    anomaly detection \cite{fanaee2016tensor} and classification \cite{phan2010tensor, zink2016tensor}.

    The research paradigm which lies at the intersection between ensemble learning and high dimensional data processing is known as \textit{multi-view learning}.
    The concept of the multi-view ensemble learning was proposed in \cite{kumar2016multi}, however, despite its intrinsic multi-way structure, the data are still considered as
    single-way entities (vectors); this does not admit natural dimensionality reduction that would take advantage of the rich latent structure present in data.
    Furthermore, the analysis is limited to finding only relevant and irrelevant features.
    Other approaches under the framework of multiple-view multiple learners has been successfully
    applied to semi-supervised and active learning \cite{zhang2010multiple, sun2011multiple},
    but still without employing multidimensional arrays, tensors, as a compact, rigorous and inherently structure-preserving model.
    The aim of this work is therefore to fill this void in the open literature and
    to introduce a novel framework which naturally incorporates physically meaningful tensor decompositions
    into ensemble learning of the intrinsically multi-way and multi-modal data.

\section{Theoretical Background} \label{sec:background}
    This paper adopts the following notation:
    a scalar is denoted by an italic lowercase letter, $x \in \R$;
    a vector by boldface lowercase letters, $\mat{x} \in \R^{I}$;
    a matrix by boldface uppercase letters, $\mat{X} \in \R^{I \times J}$;
    a tensor by underlined boldface capital letters, $\ten{X} \in \R^{I \times J \times K}$;
    a dataset of $M$ samples, $\mat{x}^m$, and the corresponding labels, $y^m$, are  designated as $\mathcal{D}:\{(\mat{x}^m, y^m)\}$;
    a classifier, $\mathcal{C}$, is denoted by $\mathcal{C}(\mathcal{D})$ during the training stage,
    whereas at the testing phase it is represented as $\mathcal{C}(\mat{x})$.

\subsection{Ensemble Learning: The Bagging Approach}
    Consider an ensemble of $N$ independent classifiers, $\mathcal{C}=\{\mathcal{C}_1,\ldots, \mathcal{C}_N\}$,
    employed for a
    binary classification problem based on a dataset $\mathcal{D}:$ $\{(\mat{x}^1, y^1), \ldots,$ $(\mat{x}^M,y^M) \} $,
    where $y^m\in\{0,1\}$ and $m = 1, \ldots, M$.
    Provided that every classifier within the ensemble, $C_n$, missclassifies the previously unseen sample, $\mat{x}^{new}$, with probability $p<0.5$,
    we can write
    \begin{equation}
    \begin{gathered}
        \mathcal{P}(\mathcal{C}) = \sum^N_{n=\lceil\frac{N}{2}\rceil} \binom{N}{n} p^n(1-p)^{N-n}\\
        \lim_{N \to \infty} \mathcal{P}(\mathcal{C}) = 0
    \end{gathered}
    \end{equation}
    where $\mathcal{P}(\cdot)$ is the probability of majority voting being incorrect. It has been proved that
    the operator $\mathcal{P}$ is monotonically decreasing with $N$ \cite{lam1997application},
    which simply implies that the more base learners participate in the majority vote the more accurate their collective decision.
    The two conditions that should be satisfied for this to hold true are:

    \noindent 1) every predictor should perform better than a random guess,

    \noindent 2) the errors of individual predictors should be uncorrelated.

    \noindent The latter condition has been the main challenge for all bagging classifiers.
    In principle, there are two ways to address this problem:
    \begin{itemize}[itemindent=0.3cm, leftmargin=0cm]
        \item Through a heterogeneous set of base learning algorithms in order to
        introduce a degree of diversity into the individual hypotheses, despite each
        being just a crude approximation of the true underlying processes that govern the data. In this way, the individual
        under-performance of a particular learning model is compensated for when all learning models are
        combined together into a unified global system;

        \item Utilising a homogeneous set of base classifiers and exposing each member
        to a subset of the training data. This data manipulation
        procedure is called \textit{resampling}, and can significantly improve the overall generalisation ability of the model.
    \end{itemize}

    However, in real world scenarios, none of these methods is capable of completely removing correlation among the hypotheses
    generated by the base classifiers even if these are applied in conjunction.

\subsection{Multilinear Algebra: Basic Definitions}
    A tensor of \textit{order} $N$ is an N-dimensional array,
    $\ten{X} \in \R^{I_1 \times I_2 \times \cdots \times I_N}$, with a particular dimension of $\ten{X}$ referred
    to as a \textit{mode}. An element of a tensor $\ten{X}$ is a scalar $x_{i_1,i_2,\dots,i_N} = \ten{X}(i_1,i_2,\dots,i_N)$ which is indexed by $N$ indices.
    A \textit{fiber} is a vector obtained by fixing all but one of the indices, e.g.
    $\ten{X}_{(:,i_2,i_3,\dots,i_N)}$ is the mode-1 fiber.
    Mode-n unfolding is the process of element mapping from a tensor to a matrix,
    e.g. $\ten{X} \rightarrow \mat{X}_{(1)} \in \R^{I_1\times I_2 I_3\cdots I_N}$ is the mode-1 unfolding which can be visualised as
    stacking the mode-1 fibers of $\ten{X}$ as column vectors of the matrix $\mat{X}_{(1)}$. A mode-n product of a tensor $\ten{X}$
    and a matrix $\mat{A}$ is therefore equivalent to
    \begin{equation}
        \ten{Y} = \ten{X} \times_n \mat{A} \quad \Leftrightarrow \quad \mat{Y}_{(n)} = \mat{A}\mat{X}_{(n)}
    \end{equation}
    The outer product of $N$ vectors results in a \textit{rank-1} tensor of order $N$, that is
    $\mat{a}_1 \circ \mat{a}_2 \circ \cdots \circ \mat{a}_n = \ten{X} \in \R^{I_1 \times I_2 \times \cdots \times I_N}$ \cite{kolda2009tensor}.

\subsection{Higher Order Singular Value Decomposition}
\begin{figure}[t]
        \center
        \includegraphics[width=0.7\linewidth]{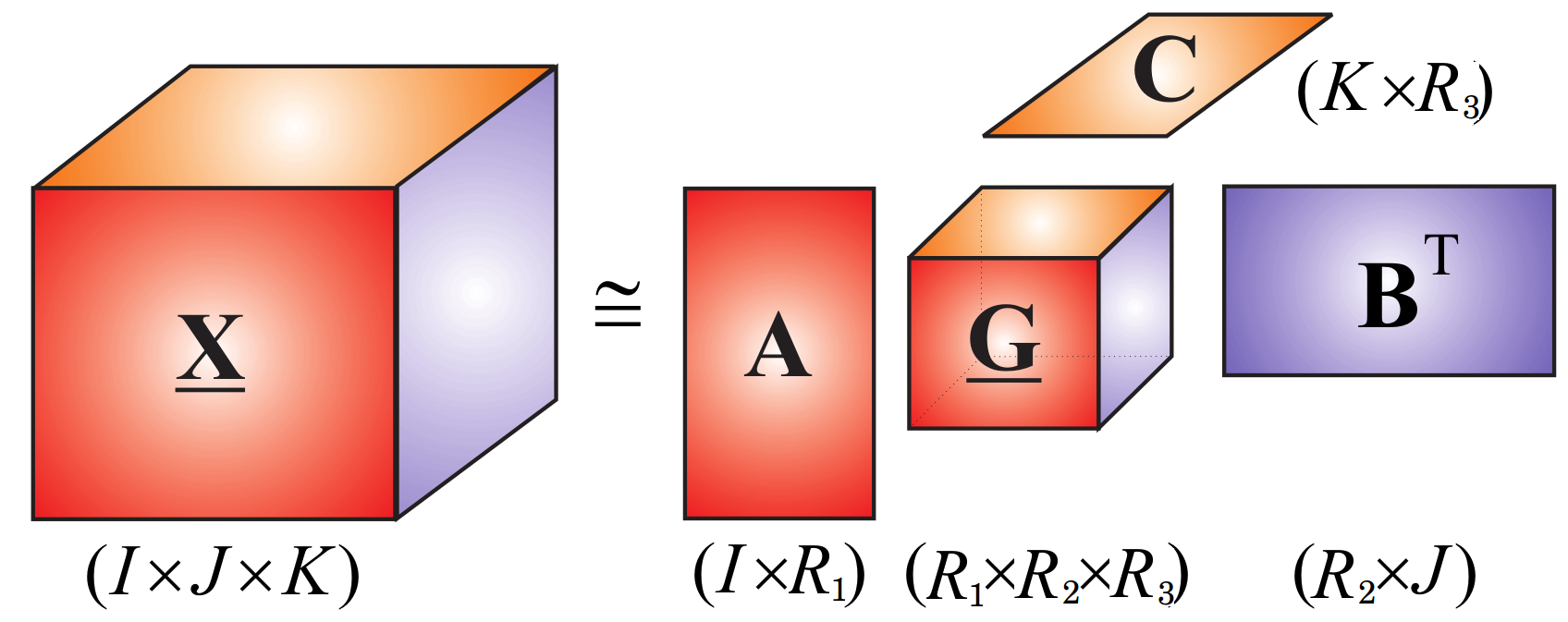}
        \caption{
            Graphical illustration of the HOSVD for a third order tensor $\ten{X}$ into a small and
            dense core tensor $\ten{G}$ and the corresponding factor matrices $\mat{A}, \mat{B}, \mat{C}$.
        }
        \label{fig:hosvd}
    \end{figure}

    A generalisation of the principal component analysis (PCA) method to multidimensional data is called the
    Higher Order Singular Value Decomposition\footnote{The HOSVD is a particular case of the Tucker Decomposition.} (HOSVD),
    which is illustrated in \cref{fig:hosvd}.
    It factorises the original tensor into the core tensor, $\ten{G}$ , and a set of factor
    matrices, $\mat{A}, \mat{B}, \mat{C}$, whereby each factor matrix corresponds to a particular mode of the original tensor, that is
    \begin{equation}
        \begin{aligned}
            \ten{X} &= \sum_{r_a=1}^{R_a} \sum_{r_b=1}^{R_b} \sum_{r_c=1}^{R_c}   \ten{X}_{r_a r_b r_c}\\
                    &= \sum_{r_a=1}^{R_a} \sum_{r_b=1}^{R_b} \sum_{r_c=1}^{R_c} g_{r_a r_b r_c} \cdot \mat{a}_{r_a} \circ \mat{b}_{r_b} \circ \mat{c}_{r_c}\\
                    &= \ten{G} \times_1 \mat{A} \times_2 \mat{B} \times_3 \mat{C} \\
                    &= \td{\ten{G};\mat{A} , \mat{B}, \mat{C}}
        \end{aligned}
        \label{eq:hosvd}
    \end{equation}
    Here, a 3-rd order tensor, $\ten{X} \in \R^{I \times J \times K}$, is factorised into a much smaller and dense
    core tensor, $\ten{G} \in \R^{R_a \times R_b \times R_c}$, and a set of orthogonal factor matrices,
    $\mat{A} \in \R^{I \times R_a}, \mat{B} \in \R^{J \times R_b},  \mat{C} \in \R^{K \times R_c}$.
    The computation of the HOSVD is straightforward and was originally proposed in \cite{de2000multilinear}.
    At first, all factor matrices are computed as left singular matrices of all possible unfoldings of a tensor, to yield
    \begin{equation}
        \begin{aligned}
            \mat{X}_{(1)} &= \mat{A} \mat{\Sigma}_1 \mat{V}_1^T\\
            \mat{X}_{(2)} &= \mat{B} \mat{\Sigma}_2 \mat{V}_2^T\\
            \mat{X}_{(3)} &= \mat{C} \mat{\Sigma}_3 \mat{V}_3^T\\
        \end{aligned}
        \label{eq:hosvd-step1}
    \end{equation}
    Then, the core tensor is computed as a projection of the original tensor on the multimodal subspaces spanned by the
    factor matrices computed at the previous step, that is
    \begin{equation}
        \ten{G}  = \ten{X} \times_1 \mat{A}^T \times_2 \mat{B}^T \times_3 \mat{C}^T
        \label{eq:hosvd-step2}
    \end{equation}
    Given the significant redundancy when a tensor $\ten{X}$ is unfolded along any mode, it is a common practice
    to use a truncated SVD in \cref{eq:hosvd-step1}, performed by keeping only the first most significant $R_a, R_b$ and $R_c$ singular vectors
    of the corresponding matrices $\mat{A}, \mat{B}$ and $\mat{C}$. For a general order-$N$ tensor, the $N$-tuple $(R_1, \ldots, R_N)$ is called
    the \textit{multi-linear rank} and provides flexibility and enables super-compression in the approximation of the original tensor.
    For our order-$3$ tensor in \cref{eq:hosvd} the multilinear rank is therefore $(R_a, R_b, R_c)$.

\section{Proposed TEL Framework} \label{sec:method}
    As described previously, current methods for generating ensembles of estimators rely on resampling the
    input data and/or using a set of different base classifiers for same data.
    Neither approach takes into account the original structure intrinsic to the data and the underlying
    mutimodal dependencies.
    To this end, we propose a natural learning framework for multi-way data,
    referred to as \textit{tensor ensemble learning} (TEL), which enables ensemble
    construction based on direct application of multilinear analysis.
    In this way, the inherently rich structure of the input samples is fully exploited which
    equips the framework with the ability to extract otherwise indistinguishable latent information from
    multidimensional data. The proposed TEL method comprises the four main stages:

    \begin{figure}[t]
        \scalebox{0.88}{
            \begin{tikzpicture}
    \tikzset{
        every node/.style={node distance=2cm,
                           align=center,
                           minimum height=1cm,
                           minimum width=\linewidth,
                           inner color=white
        },
        block/.style={rectangle,
                      draw=black,
                      outer color=SeaGreen3,
                      rounded corners
        },
        stage/.style={rectangle,
                      draw=white,
                      outer color=white,
                      minimum width=0.01\linewidth,
                      inner sep=0.1pt,
                      anchor=east,
                      rounded corners
        },
        decomposition/.style={outer color=blue!30
        },
        dataset/.style={minimum width=2cm
        },
        classifier/.style={circle,
                           minimum width=2cm,
                           draw,
                           outer color=red!30
        },
        circle dotted/.style={dash pattern=on 0.5pt off 10pt, line cap=round,line width = 2pt},
        arrow/.style = {-{Stealth[scale=1]}, rounded corners, very thick},
    }

    \newcommand{\printDataset}{
        Training data \\ $ \mathcal{D}$ : $\Big\{ (\ten{X}^m, y^m) \Big\}, \quad \ten{X}^m \in \R^{I \times J \times K} \text{ \ for } m = 1,\ldots,M$
    }
    \newcommand{\printDecomp}{
        Tensor factorisation of each sample from $\mathcal{D}$ \\ [1mm]
        $\ten{X}^m = \td{\ten{G}^m;\mat{A}^m , \mat{B}^m, \mat{C}^m}, \quad m = 1,\ldots,M$ \\ [2mm]
        $\mat{A}^m = \Big[ \mat{a}^m_1 \cdots \mat{a}^m_{R_a} \Big];$
        $\mat{B}^m = \Big[ \mat{b}^m_1 \cdots \mat{b}^m_{R_b} \Big];$
        $\mat{C}^m = \Big[ \mat{c}^m_1 \cdots \mat{c}^m_{R_c} \Big];$
    }
    \newcommand{\printRegroup}{
        Regrouping of the factor vectors into separate datasets\\ [1mm]
        $\mathcal{D}^A_1$:$\Big\{ (\mat{a}^m_1, y^m)\Big\}; \cdots ; \mathcal{D}^A_{R_a}$:$\Big\{ (\mat{a}^m_{R_a}, y^m)\Big\}, \quad m = 1,\ldots,M$ \\
        $\mathcal{D}^B_1$:$\Big\{ (\mat{b}^m_1, y^m)\Big\}; \cdots ; \mathcal{D}^B_{R_b}$:$\Big\{ (\mat{b}^m_{R_b}, y^m)\Big\}, \quad m = 1,\ldots,M$ \\
        $\mathcal{D}^C_1$:$\Big\{ (\mat{c}^m_1, y^m)\Big\}; \cdots ; \mathcal{D}^C_{R_c}$:$\Big\{ (\mat{c}^m_{R_c}, y^m)\Big\}, \quad m = 1,\ldots,M$ \\
    }
    \newcommand{\printFirstClassifier}{
        Train\\
        Classifier\\ [1mm]
        $\mathcal{C}_{1}(\mathcal{D}^A_1)$
    }
    \newcommand{\printLastClassifier}{
        Train\\
        classifier\\ [1mm]
        $\mathcal{C}_{N}(\mathcal{D}^C_{R_c})$
    }
    \newcommand{\printVote}{
        For the $\ten{X}^{new} = \td{ \ten{G}^{new}; \mat{A}^{new}, \mat{B}^{new}, \mat{C}^{new} }$, assign label $y^{new}$\\ [1mm]
        based on majority vote of $\Big\{ \mathcal{C}_{1}(\mat{a}^{new}_1), \ldots, \mathcal{C}_{N}(\mat{c}^{new}_{R_c}) \Big\}$
    }
    \newcommand{\printStage}[1]{
        \textbf{\underline{Stage: #1}}
    }


\node (dataset) [block] {\printDataset};
\node (decomposition) [block, decomposition, below of=dataset, node distance=2.5cm] {\printDecomp};
\node (regroup) [block, decomposition, below of=decomposition, node distance=3.5cm] {\printRegroup};
\node (clfA) [classifier, anchor=west]  at ($(regroup.south west)-(0,1.5cm)$) {\printFirstClassifier};
\node (clfC) [classifier, anchor=east]  at ($(regroup.south east)-(0,1.5cm)$) {\printLastClassifier};
\node (dummy) [minimum width=0.1cm] at ($(clfA)!0.5!(clfC)$) {};
\node (vote) [block, below of=dummy, outer color=red!30,node distance=2.5cm] {\printVote};
\node (s) [block, stage] at ($(decomposition.west) + (-0.1cm,0)$) {\printStage{1}};
\node (s) [block, stage] at ($(regroup.west) + (-0.1cm,0)$) {\printStage{2}};
\node (s) [block, stage] at ($(clfA.west) + (-0.1cm,0)$) {\printStage{3}};
\node (s) [block, stage] at ($(vote.west) + (-0.1cm,0)$) {\printStage{4}};

\draw[arrow]  (dataset) -- (decomposition);
\draw[arrow]  (decomposition) -- (regroup);
\draw[arrow]  (regroup) -- (clfA);
\draw[arrow]  (regroup) -- (clfC);
\draw[arrow, dashed]  (clfA) -- (vote);
\draw[arrow, dashed]  (clfC) -- (vote);
\draw[circle dotted] (clfA) -- node[anchor=south, minimum width=1cm] {Total number of classifiers \\$N=R_a+R_b+R_c$} (clfC);

\end{tikzpicture}
        }
        \caption{
            Tensor ensemble learning through the HOSVD with multi-linear rank $(R_a, R_b, R_c)$. All factor vectors of training samples,
            $\ten{X}^m$, are reorganised into separate datasets, thus only requiring to train $N = R_a + R_b + R_c$ classifiers.
            At the final stage, a majority vote is used  to classify a new unlabelled sample, $\ten{X}^{new}$,
            achieved by aggregation of the individually learned knowledge
            about training data extrapolated to extracted latent components $\mat{a}^{new}_1, \dots, \mat{c}^{new}_{R_c}$.
        }
        \label{fig:telvi_concept}
    \end{figure}
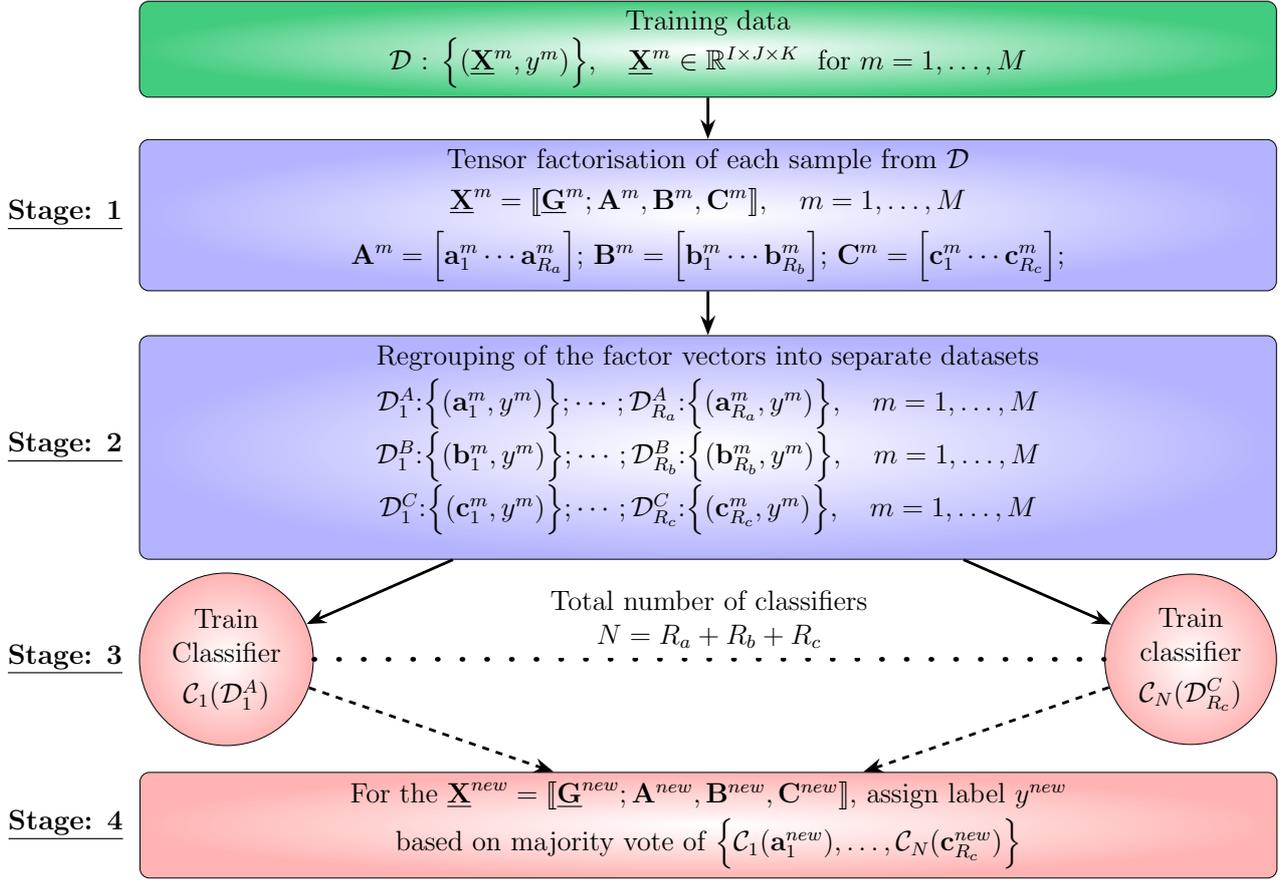

    \begin{itemize}[itemindent=2cm, leftmargin=0cm]
        \item[\textbf{Stage 1.}] Apply a tensor decomposition to each multidimensional sample from the training set in order to find latent components.
        Tensor decompositions that can be utilised at this stage include
        Canonical Polyadic Decomposition (CPD) \cite{kiers2001three},
        Higher Order Singular Value Decomposition (HOSVD) \cite{de2000multilinear},
        Block Term Decomposition (BTD) \cite{de2008decompositions} and the
        Tensor Train Decomposition (TT) \cite{oseledets2011tensor, cichocki2016tensor, cichocki2017tensor}.

        \item[\textbf{Stage 2.}] Perform reorganisation of the components extracted  in Stage 1 in order to generate multiple new datasets,
        each containing an incomplete information about the original sample.
        Given the different notions of existing tensor decompositions,
        there are multiple ways to accomplish this step.

        \item[\textbf{Stage 3.}] Train an ensemble of base learners in order to generate a series of independent hypotheses about
        the underlying processes that govern the datasets constructed in Stage 2.

        \item[\textbf{Stage 4.}] When facing a previously unseen data sample, aggregate the knowledge about its latent components based on
        the hypotheses from Stage 3. This can be achieved using any basic method, such as the weighted majority vote,
        or through more advanced machine learning algorithms.

    \end{itemize}

    Having established such a general framework of the tensor ensemble learning, we next
    demonstrate a successful implementation (depicted in \cref{fig:telvi_concept}) of the proposed concept
    in the context of supervised multi-class classification.
    For simplicity of presentation and ease of illustration we restrict ourselves to tensors of order $3$.
    The solution in Section 3.1 can be straightforwardly generalised to tensors of any arbitrary order.

\subsection{Tensor Ensemble Learning for Classification}
    Consider the problem of training a classifier using a dataset, $\mathcal{D}$, of $M$ labelled samples, which are naturally multidimensional,
    and come from $K$ distinct categories, that is
    \begin{align}
        \mathcal{D}:\{(\ten{X}^m, y^m)\}, \quad m = 1,\ldots, M
    \end{align}
    Here, $y^m \in [1,K]$ is a label associated with the training sample $\ten{X}^m \in \R^{I \times J \times K}$.
    Recall that an important assumption behind the success of ensemble learning is that the hypotheses from each base learner are independent.
    This can be enforced by providing uncorrelated datasets to every base learner.
    A tensor decomposition that guarantees such uncorrelated factors is the HOSVD.
    Therefore, an implementation of Stage 2 of the TEL framework using HOSVD with the multi-linear rank $(R_a, R_b, R_c)$
    yields the following representation of the original data set
    \begin{equation}
    \begin{aligned}
        \mathcal{D}':\Big\{ \Big( \td{\ten{G}^m;\mat{A}^m , \mat{B}^m, \mat{C}^m}, y^m \Big) \Big\}, \quad m = 1,\ldots, M
    \end{aligned}
    \end{equation}
    Here, the orthonormal factor matrices $\mat{A}^m \in \R^{I \times R_a} , \mat{B}^m \in \R^{J \times R_b}, \mat{C}^m \in \R^{K \times R_c}$
    are uniquely defined by the corresponding column vectors $\mat{a}^m_{r_a}, \mat{b}^m_{r_b}$ and $\mat{c}^m_{r_c}$,
    where $r_a \in [1, R_a]; r_b \in [1, R_b]$ and $r_c \in [1, R_c]$.
    Thus, if a reorganisation is performed such that these factor vectors are considered by the base estimators independently,
    this would provide an ensemble learning framework which is practically feasible and yields enhanced
    performance after aggregation of the individual hypotheses.
    A detailed illustration of such reorganisation is given in Stage 2 of \cref{fig:telvi_concept}.
    Consequently, there are $N=R_a+R_b+R_c$ datasets which are separately fed into $N$ base learners at the next stage.
    In this particular implementation of the TEL, the final stage is accomplished using
    a simple majority vote but other aggregation methods could be equally utilised.

    To summarise, the proposed Tensor Ensemble Learning -- Vector Independent (TELVI) estimator is a machine learning model that belonds to the TEL framework,
    whereby the resampling of the factor vectors is performed as depicted in Stage 2 of \cref{fig:telvi_concept}.

\section{Simulations and Analysis} \label{sec:simulation_results}

    The proposed approach was employed for object classification from images
    within the benchmark ETH-80 dataset \cite{leibe2003analyzing}. This dataset consists of
    3280 colour images ($128 \times 128$ pixels) from 8 categories: apple, car, cow, cup, dog, horse, pear, tomato.
    Each category contains 10 different objects with 41 views per object, spaced equally over the viewing hemisphere.
    Observe that the samples of ETH-80 dataset are inherently multidimensional since any colour (RGB) image can be
    represented as a three-dimensional array $pixel\_X \times pixel\_Y \times 3$.
    For the simulations we used \texttt{Scikit-Learn} \cite{scikit-learn}, a highly optimised
    library of classical machine learning algorithms, while the computation of the HOSVD and implementation of the
    TELVI classifier \footnote{Available within Jupyter notebooks and an alpha version of HOTTBOX on request.}
    were implemented
    using our own recently developed Python software package for multidimensional data, \texttt{HOTTBOX},
    publicly available at \cite{hottbox}.
    To assess performance of the proposed and classical approaches, we used a ratio of correctly categorised images to the total number of samples in the test set.

    \begin{figure}[t]
        \center
        \includegraphics[width=0.7\linewidth,trim={0 2mm 0 2mm}, clip]{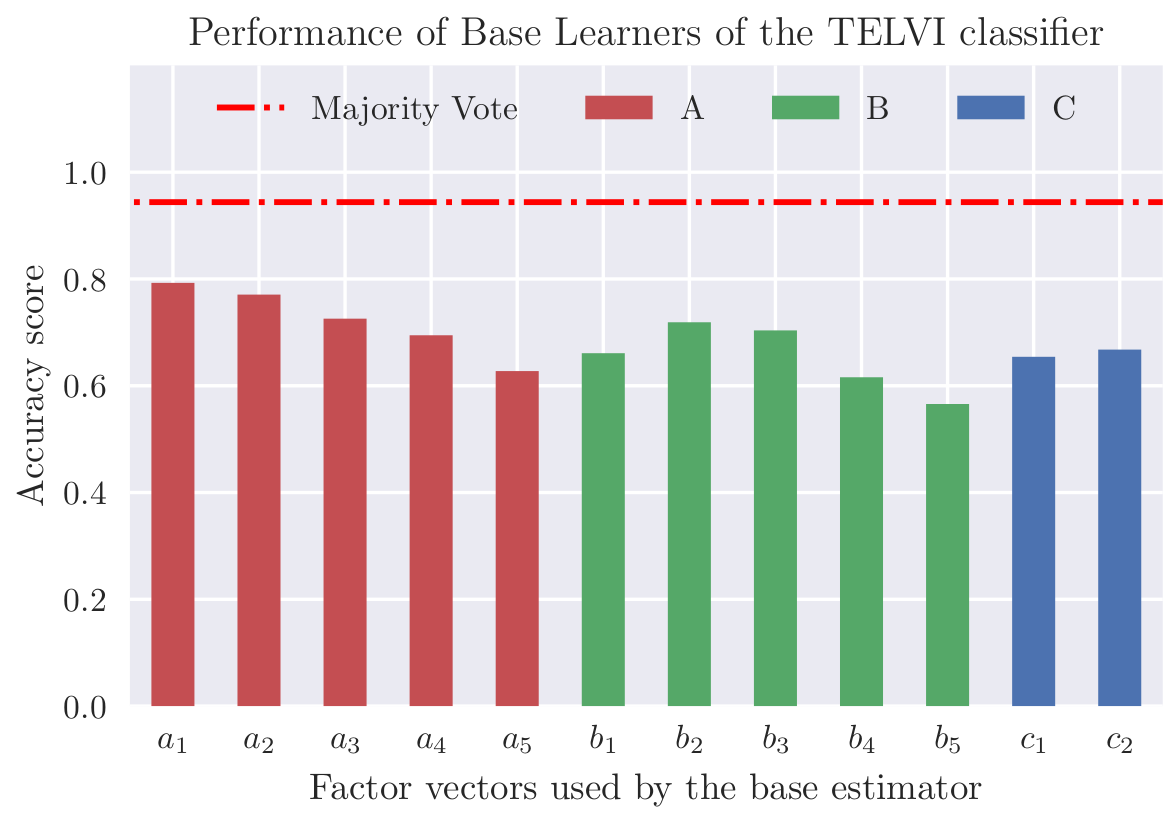}
        \caption{
            Performance of the employed base classifiers which were individually trained and used for prediction of
            a particular latent component from factor matrices $\mat{A}, \mat{B}$ and $ \mat{C}$, computed through HOSVD.
        }
        \label{fig:base_clf_performance}
    \end{figure}

    Prior to comparing the overall performances of the TELVI and Bagging approaches,
    classification accuracy scores of each chosen base classifier within the TELVI were computed
    in order to establish the behavioural correspondence between this implementation of TEL framework
    and classical ensemble learning. The classification pipeline for the TELVI classifier begins with
    splitting the images from the ETH-80 dataset into the training and test sets.
    This was followed by extracting the latent components from each image individually with the use of
    the HOSVD algorithm for the multi-linear rank $(5,5,2)$. The latter should be treated as the hyper-parameter of the
    TELVI classifier. In our simulations, its values were chosen so as to give
    the optimal ratio between the compression rate and the approximation
    error of a particular tensor decomposition across all images.
    Next, the factor components of images from the training set were regrouped
    (as illustrated in Stage 2 of \cref{fig:telvi_concept})
    into 12 new training datasets, thus requiring a total of 12
    independent base learners to be employed.
    After each base classifier had been trained, the TELVI classifier was ready to
    assign labels for the test samples that would undergo the same ``preprocessing'' steps
    (decomposing a test sample through HOSVD with multi-linear rank
    $(5,5,2)$ and regrouping obtained components).
    Finally, a label was assigned based on the majority vote over a list of labels produced
    by the base classifiers.

    \cref{fig:base_clf_performance} shows the individual test performance of
    each base classifier (Support Vector Machine with the polynomial kernel) trained with 50\% of the original dataset.
    Observe that none of the base classifiers exhibited strong performance
    on the training set, with the highest reaching 80\%. However, when the individual
    classifiers were combined through a majority vote,
    the accuracy score of 94\% was achieved. This is significantly higher than for any of the base estimators
    and provides a conclusive evidence that the TELVI classifier operates in a manner similar to classic ensemble learning approaches.

    \begin{figure}[t]
        \includegraphics[width=\linewidth]{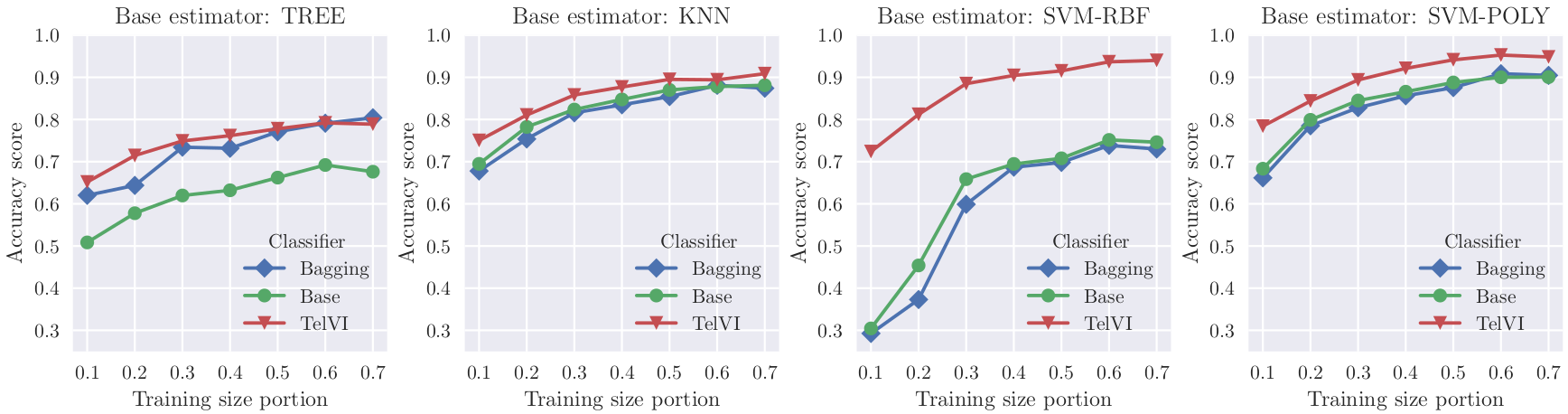}
        \caption{
            Comparison of the overall test performance for Bagging and TELVI approaches on the ETH-80 dataset, illustrated
            as a function of the proportion of training size proportion and across a variety of classification algorithms used as the base estimators.
        }
        \label{fig:overall-comparion}
    \end{figure}

    Next, the proposed method was rigorously assessed against the Bagging approach through four experiments.
    The images from the ETH-80 dataset were randomly split into the training and test data
    with the size of training proportion ranging from 10\% to 70\% of all available samples.
    In each experiment, all base learners of both TELVI and Bagging classifiers used the same machine learning algorithm:
    Support Vector Machines with both the Gaussian (SVM-RBF) and polynomial (SVM-POLY) kernel, Decision Trees (TREE) and K-nearest neighbours (KNN).
    Hyperparameters were tuned based on an exhaustive grid search with the 5-fold cross validation of the training data.
    The classification pipeline for the TELVI remained the same as described above. Recall that
    Bagging approach requires to start with the vectorisation of all samples from the training and test data.
    Given the large number of features, dimensionality reduction though Principal Component Analysis (PCA) was used
    to reduce computational complexity during the training stage.
    For fair comparison, the Bagging classifier also utilised 12 base learners that were combined
    by virtue of the majority vote.

    The results of this set of experiments are presented in \cref{fig:overall-comparion}.
    Observe that TEL approach achieved significantly increased classification rates, independent of the nature
    of the particular base estimator at hand, with the highest scores reaching 95\% for SVM-POLY,
    while Bagging performed at around the same level as the base classifier,
    except for the Decision Tree.
    This conclusively demonstrates that, even when resampling with replacement did not provide the desired increase in the performance,
    exploiting the inherent structure within the data is a promising way to create a series of datasets \notshow{with enhanced discrimination ability,} as a data manipulation
    part of ensemble learning.

\section{Conclusion}
    We have introduce tensor ensemble learning (TEL) as a novel framework for generating ensembles
    of base estimators for multidimensional data. This has been achieved by virtue of tensor decompositions
    thus making TEL highly parallelisable and suitable for large-scale problems.
    As one of the possible implementations of TEL, we have presented the TELVI algorithm
    that has been derived in accordance with the definition of classical ensemble learning.
    The conducted simulations have shown that when compared with the popular approach of \textit{bootstrap aggregating},
    TELVI is capable of improving test performance of a multi-class image classification task.
    This success is partially due to its ability to obtain uncorrelated surrogate datasets that are generated by HOSVD and
    partially owing to the dimensionality reduction inherent to tensor decompositions which takes into account the original structure of the data.

\newpage
\bibliographystyle{IEEE}
\bibliography{refs}

\begin{thebibliography}{10}

\bibitem{dietterich2000ensemble}
T.~G. Dietterich,
\newblock ``Ensemble methods in machine learning,''
\newblock {\em In Proceedings of International Workshop on Multiple Classifier
  Systems}, pp. 1--15, 2000.

\bibitem{dietterich2000experimental}
T.~G Dietterich,
\newblock ``An experimental comparison of three methods for constructing
  ensembles of decision trees: Bagging, boosting, and randomization,''
\newblock {\em Machine Learning}, vol. 40, no. 2, pp. 139--157, 2000.

\bibitem{opitz1999popular}
D.~Opitz and R.~Maclin,
\newblock ``Popular ensemble methods: An empirical study,''
\newblock {\em Journal of Artificial Intelligence Research}, vol. 11, pp.
  169--198, 1999.

\bibitem{dvzeroski2004combining}
S.~D{\v{z}}eroski and B.~{\v{Z}}enko,
\newblock ``Is combining classifiers with stacking better than selecting the
  best one?,''
\newblock {\em Machine Learning}, vol. 54, no. 3, pp. 255--273, 2004.

\bibitem{koren2009bellkor}
Y.~Koren,
\newblock ``{The BellKor solution to the Netflix grand prize},''
\newblock {\em Netflix Prize Documentation}, vol. 81, pp. 1--10, 2009.

\bibitem{schapire1990strength}
R.~E. Schapire,
\newblock ``The strength of weak learnability,''
\newblock {\em Machine Learning}, vol. 5, no. 2, pp. 197--227, 1990.

\bibitem{freund1997decision}
Y.~Freund and R.~E. Schapire,
\newblock ``A decision-theoretic generalization of on-line learning and an
  application to boosting,''
\newblock {\em Journal of Computer and System Sciences}, vol. 55, no. 1, pp.
  119--139, 1997.

\bibitem{wolpert1992stacked}
D.~H. Wolpert,
\newblock ``Stacked generalization,''
\newblock {\em Neural Networks}, vol. 5, no. 2, pp. 241--259, 1992.

\bibitem{breiman1996bagging}
L.~Breiman,
\newblock ``Bagging predictors,''
\newblock {\em Machine Learning}, vol. 24, no. 2, pp. 123--140, 1996.

\bibitem{liaw2002classification}
A.~Liaw, M.~Wiener, and \textit{et al}.,
\newblock ``Classification and regression by {randomForest},''
\newblock {\em R News}, vol. 2, no. 3, pp. 18--22, 2002.

\bibitem{cichocki2015tensor}
A.~Cichocki, D.~P. Mandic, L.~De~Lathauwer, G.~Zhou, Q.~Zhao, C.~Caiafa, and
  H.~A. Phan,
\newblock ``Tensor decompositions for signal processing applications: From
  two-way to multiway component analysis,''
\newblock {\em IEEE Signal Processing Magazine}, vol. 32, no. 2, pp. 145--163,
  2015.

\bibitem{wong2015joint}
W.~K. Wong, Z.~Lai, Y.~Xu, J.~Wen, and C.~P. Ho,
\newblock ``Joint tensor feature analysis for visual object recognition,''
\newblock {\em IEEE Transactions on Cybernetics}, vol. 45, no. 11, pp.
  2425--2436, 2015.

\bibitem{kisil2018common}
I.~Kisil, G.~G. Calvi, A.~Cichocki, and D.~P. Mandic,
\newblock ``Common and individual feature extraction using tensor
  decompositions: A remedy for the curse of dimensionality?,''
\newblock {\em In Proceedings of the IEEE International Conference on
  Acoustics, Speech and Signal Processing (ICASSP)}, pp. 6299--6303, 2018.

\bibitem{acar2013understanding}
E.~Acar, M.~A. Rasmussen, F.~Savorani, T.~N{\ae}s, and R.~Bro,
\newblock ``Understanding data fusion within the framework of coupled matrix
  and tensor factorizations,''
\newblock {\em Chemometrics and Intelligent Laboratory Systems}, vol. 129, pp.
  53--63, 2013.

\bibitem{calvi2018sum}
G.~G. Calvi, I.~Kisil, and D.~P. Mandic,
\newblock ``Feature fusion via tensor network summation,''
\newblock {\em In Proceedings of the 26th European Signal Processing Conference
  (EUSIPCO)}, p. TBA, 2018.

\bibitem{fanaee2016tensor}
H.~Fanaee-T and J.~Gama,
\newblock ``Tensor-based anomaly detection: An interdisciplinary survey,''
\newblock {\em Knowledge-Based Systems}, vol. 98, pp. 130--147, 2016.

\bibitem{phan2010tensor}
A.~Phan and A.~Cichocki,
\newblock ``Tensor decompositions for feature extraction and classification of
  high dimensional datasets,''
\newblock {\em Nonlinear Theory and its Applications}, vol. 1, no. 1, pp.
  37--68, 2010.

\bibitem{zink2016tensor}
R.~Zink, B.~Hunyadi, S.~Van~Huffel, and M.~De~Vos,
\newblock ``Tensor-based classification of an auditory mobile {{BCI}} without a
  subject-specific calibration phase,''
\newblock {\em Journal of Neural Engineering}, vol. 13, no. 2, pp. 1--10, 2016.

\bibitem{kumar2016multi}
V.~Kumar and S.~Minz,
\newblock ``Multi-view ensemble learning: An optimal feature set partitioning
  for high-dimensional data classification,''
\newblock {\em Knowledge and Information Systems}, vol. 49, no. 1, pp. 1--59,
  2016.

\bibitem{zhang2010multiple}
Q.~Zhang and S.~Sun,
\newblock ``Multiple-view multiple-learner active learning,''
\newblock {\em Pattern Recognition}, vol. 43, no. 9, pp. 3113--3119, 2010.

\bibitem{sun2011multiple}
S.~Sun and Q.~Zhang,
\newblock ``Multiple-view multiple-learner semi-supervised learning,''
\newblock {\em Neural Processing Letters}, vol. 34, no. 3, pp. 229, 2011.

\bibitem{lam1997application}
L.~Lam and S.~Suen,
\newblock ``Application of majority voting to pattern recognition: An analysis
  of its behaviour and performance,''
\newblock {\em IEEE Transactions on Systems, Man, and Cybernetics -- Part A:
  Systems and Humans}, vol. 27, no. 5, pp. 553--568, 1997.

\bibitem{kolda2009tensor}
T.~G. Kolda and B.~W. Bader,
\newblock ``Tensor decompositions and applications,''
\newblock {\em SIAM Review}, vol. 51, no. 3, pp. 455--500, 2009.

\bibitem{de2000multilinear}
L.~De~Lathauwer, B.~De~Moor, and J.~Vandewalle,
\newblock ``A multilinear singular value decomposition,''
\newblock {\em SIAM Journal on Matrix Analysis and Applications}, vol. 21, no.
  4, pp. 1253--1278, 2000.

\bibitem{kiers2001three}
H.~A. Kiers and I.~V. Mechelen,
\newblock ``Three-way component analysis: Principles and illustrative
  application,''
\newblock {\em Psychological Methods}, vol. 6, no. 1, pp. 84--110, 2001.

\bibitem{de2008decompositions}
L.~De~Lathauwer,
\newblock ``Decompositions of a higher-order tensor in block terms. {{Part}}
  {{II}}: Definitions and uniqueness,''
\newblock {\em SIAM Journal on Matrix Analysis and Applications}, vol. 30, no.
  3, pp. 1033--1066, 2008.

\bibitem{oseledets2011tensor}
I.~Oseledets,
\newblock ``Tensor-train decomposition,''
\newblock {\em SIAM Journal on Scientific Computing}, vol. 33, no. 5, pp.
  2295--2317, 2011.

\bibitem{cichocki2016tensor}
A.~Cichocki, N.~Lee, I.~Oseledets, A.~H. Phan, Q.~Zhao, and D.~P. Mandic,
\newblock ``Tensor networks for dimensionality reduction and large-scale
  optimization. {{Part}} 1: {{Low-rank}} tensor decompositions,''
\newblock {\em Foundations and Trends in Machine Learning}, vol. 9, no. 4-5,
  pp. 249--429, 2016.

\bibitem{cichocki2017tensor}
A.~Cichocki, A.~H. Phan, Q.~Zhao, N.~Lee, I.~Oseledets, M.~Sugiyama, and D.~P.
  Mandic,
\newblock ``Tensor networks for dimensionality reduction and large-scale
  optimization. {{Part}} 2: {{Applications}} and future perspectives,''
\newblock {\em Foundations and Trends in Machine Learning}, vol. 9, no. 6, pp.
  431--673, 2017.

\bibitem{leibe2003analyzing}
B.~Leibe and B.~Schiele,
\newblock ``Analyzing appearance and contour based methods for object
  categorization,''
\newblock {\em In Proceedings of the IEEE Computer Society Conference on
  Computer Vision and Pattern Recognition (CVPR)}, vol. 2, pp. 409--415, 2003.

\bibitem{scikit-learn}
F.~Pedregosa, G.~Varoquaux, A.~Gramfort, V.~Michel, B.~Thirion, O.~Grisel,
  M.~Blondel, P.~Prettenhofer, R.~Weiss, V.~Dubourg, J.~Vanderplas, A.~Passos,
  D.~Cournapeau, M.~Brucher, M.~Perrot, and E.~Duchesnay,
\newblock ``{Scikit-learn: Machine learning in python },''
\newblock {\em Journal of Machine Learning Research}, vol. 12, pp. 2825--2830,
  2011.

\bibitem{hottbox}
I.~Kisil, A.~Moniri, G.~G. Calvi, B.~Scalzo~Dees, and D.~P. Mandic,
\newblock ``{HOTTBOX: Higher Order Tensors ToolBOX},'' \ \ \ \
  \url{https://github.com/hottbox}.

\end{thebibliography}

\end{document}